\documentclass[usenatbib,letterpaper]{mn2e}
\usepackage{epsfig}
\usepackage{amsmath}
\title[Galaxy ecology in SDSS and 2dFGRS]{Galaxy ecology: groups and low-density environments in
  the SDSS and 2dFGRS}
\author[Balogh \etal]{Michael Balogh$^{1}$, Vince Eke$^{1}$, Chris  Miller$^{2}$, Ian Lewis$^{3}$, Richard Bower$^{1}$,  \newauthor
Warrick Couch$^{4}$,   Robert Nichol$^{2}$,  Joss Bland-Hawthorn$^{5}$, Ivan K. Baldry$^{6}$, \newauthor
Carlton Baugh$^{1}$, Terry Bridges$^5$, Russell Cannon$^5$, Shaun Cole$^1$, Matthew Colless$^7$,  \newauthor
Chris Collins$^8$, Nicholas Cross$^{7,9}$, Gavin Dalton$^4$, Roberto De Propris$^{4}$, \newauthor
Simon P. Driver$^{9}$, George Efstathiou$^{10}$, Richard S. Ellis$^{11}$,  Carlos S. Frenk$^1$, \newauthor
Karl Glazebrook$^6$, Percy Gomez$^{2}$, Alex Gray$^{12}$, Edward Hawkins$^{13}$, Carole Jackson$^{7}$, \newauthor
Ofer Lahav$^{10}$, Stuart Lumsden$^{14}$, Steve Maddox$^{13}$, Darren Madgwick$^{10}$, Peder Norberg$^{15}$, \newauthor
John A. Peacock$^{16}$, Will Percival$^{16}$, Bruce A. Peterson$^7$, Will Sutherland$^{16}$, \newauthor
Keith Taylor$^{11}$
\\
$^{1}$Department of Physics, University of Durham, South Road, Durham DH1 3LE, UK\\
$^{2}$Department of Physics, Carnegie Mellon University, 5000 Forbes Avenue, Pittsburgh, PA 15213, USA\\
$^{3}$Astrophysics, Nuclear and Astrophysics Laboratory, Keble Road, Oxford OX1 3RH, UK\\
$^{4}$School of Physics, University of New South Wales, Sydney 2052, Australia\\
$^{5}$Anglo-Australian Observatory, P.O. Box 296, Epping, NSW 1710, Australia \\
$^{6}$Department of Physics and Astronomy, Johns Hopkins University, 3400 North Charles Street, Baltimore, MD 21218-2686 USA\\
$^7$Research School of Astronomy \& Astrophysics, The Australian National University, Weston Creek, ACT 2611, Australia \\
$^8$Astrophysics Research Institute, Liverpool John Moores University, Twelve Quays House, Egerton Wharf, Birkenhead, L14 1LD, UK \\
$^{9}$School of Physics and Astronomy, University of St. Andrews, North Haugh, St Andrews, Fife KY16 9SS, UK\\
$^{10}$Institute of Astronomy, University of Cambridge, Madingley Road, Cambridge\\ 
$^{11}$California Institute of Technology, Pasadena, CA, 91125-2400, U.S.A.\\
$^{12}$Department of Computer Science, Carnegie Mellon University, 5000 Forbes Avenue, Pittsburgh, PA 15213, USA\\
$^{13}$School of Physics and Astronomy, University of Nottingham, University Park, Nottingham, NG7 2RD, UK \\
$^{14}$Department of Physics \& Astronomy, E C Stoner Building, Leeds LS2 9JT, UK \\
$^{15}$ETHZ Institut f\"ur Astronomie, ETH H\"onggerberg, CH-8093,    Z\"urich, Switzerland \\
$^{16}$Institute for Astronomy, University of Edinburgh, Royal Observatory, Edinburgh EH9 3HJ, UK \\
}
\date{\today}
\def\etal{{ et al.\thinspace}}
\def\gtrsim{\mathrel{\raise0.35ex\hbox{$\scriptstyle >$}\kern-0.6em
\lower0.40ex\hbox{{$\scriptstyle \sim$}}}}
\def\lesssim{\mathrel{\raise0.35ex\hbox{$\scriptstyle <$}\kern-0.6em
\lower0.40ex\hbox{{$\scriptstyle \sim$}}}}

\def\kmsmpc{{\,\rm km\,s^{-1}Mpc^{-1}}} 
\def\ewha{$W_{0}(H\alpha)$}
\def\haew{$W_{0}(H\alpha)$}
\def \75th{75${\rm th}$}
\def \25th{25${\rm th}$}
 \def\kms{{\,\rm km\,s^{-1}}}
\begin{document} 
\maketitle
\begin{abstract}We analyse the observed correlation between galaxy
  environment and 
H$\alpha$ emission line strength, using volume-limited samples and
group catalogues of 24968 galaxies at
$0.05<z<0.095$, drawn from the 2dF
Galaxy Redshift Survey ($M_{b_J}<-19.5$) and the Sloan Digital Sky
Survey ($M_r<-20.6$).  We characterise the environment by 1)
$\Sigma_5$, the surface number density of galaxies determined
by the projected distance to the 5$^{\rm th}$ nearest neighbour; and
2) $\rho_{1.1}$ and $\rho_{5.5}$, three-dimensional density
estimates obtained by convolving the galaxy distribution with Gaussian
kernels of dispersion 1.1 Mpc and 5.5 Mpc, respectively.  We find that
star-forming and quiescent galaxies form two distinct populations, as
characterised by their H$\alpha$ equivalent width, \ewha.  The relative
numbers of star-forming and quiescent galaxies varies strongly and
continuously with local density.  However, the distribution of \ewha\
amongst the star-forming population is independent of environment.
The fraction of star-forming galaxies shows strong sensitivity to the
density on large scales, $\rho_{5.5}$, which is likely
independent of the trend with local density, $\rho_{1.1}$.
We use two differently-selected group catalogues to demonstrate that the
  correlation with galaxy density is approximately independent of group
  velocity dispersion, for $\sigma=200$--$1000\kms$.  
Even in the lowest density
environments, no more than $\sim 70$ per cent of galaxies show
significant H$\alpha$ emission.   
 Based on these results, we
  conclude that 
the present-day correlation between star formation rate and environment
is a result of
  short-timescale mechanisms that take place preferentially at high
  redshift,  such as starbursts induced by galaxy-galaxy   interactions.
\end{abstract}
\section{Introduction}
The galaxy population today can be effectively described as a
combination of two distinct types.  The first are red,
morphologically early-type galaxies with little or no current star
formation; the remainder are blue, late-type galaxies with active
star formation.  This segregation has been known for a long time;
however the superb data from the Sloan Digital Sky Survey has revealed
how surprisingly distinct these two populations are, at least in terms
of their colours \citep{Strateva01_short,Baldry03}.  This segregation is
known to be strongly mass-dependent, with the most massive galaxies
being predominantly red, early-types \citep{Kauffmann-SDSS1_short}.  
It is currently unclear whether galaxy evolution at the present day
consists of galaxies changing from one type to another in a short
time, or of evolution in properties within a given class.  

Direct evidence of galaxy evolution comes from observations of
galaxies at different redshifts;  this shows that, in the past,  the average star
formation rate (SFR) was much higher \citep{L96,madau,Wilson+02}, and
the typical star-forming galaxy was more luminous
\citep{Cowie+99}.  Given the near-constancy of the SFR in the Milky 
Way \citep[e.g.][]{R-P}, it seems likely that this downsizing
effect is the main characteristic of
the decline in global star formation: the most massive galaxies have
recently stopped forming stars altogether, while less massive galaxies
continue unhindered.

A more indirect form of evolution is observed as the
change in galaxy populations as a function of their environment at a
given epoch.  In particular, galaxies in dense environments
(i.e. clusters) tend to have early-type morphologies
\citep[e.g.][]{Dressler,DML,Treu03} and low SFRs
\citep{B+97,B+98,PSG,P+99}.  One interpretation of this trend has been that the
cluster environment causes galaxies to transform their properties as
they move, pulled by gravity, from low density regions into the cluster
centre.  However, recent work \citep[e.g.][]{kap2,OHely,Kodama_cl0939}, especially that based on  the
2dF Galaxy Redshift Survey \citep[2dFGRS, ][hereafter Paper~I]{MS03,2dF-sfr_short} and Sloan
Digital Sky Survey
\citep[SDSS, ][hereafter Paper~II]{Sloan_sfr_short} has shown that the correlation between
galaxy type and local density extends to very low densities, well
beyond the region where the cluster is expected to have much influence.
These works showed that there is a smooth
dependence of SFR on local galaxy density, and identified a ``critical''
surface density of 1 Mpc$^{-2}$, where SFR correlations with environment
first occur.  
This critical density is quite low, and corresponds to regions well outside the virialised cluster
region; this suggests that galaxies may be pre-processed in groups,
before they end up in clusters.  This had been anticipated \citep[e.g.][]{ZM98,ZM00}, since
galaxy-galaxy interactions are known to induce star formation
\citep[e.g.][]{Barton,Lambas} and such 
interactions should be common in groups.  There is evidence
that these interactions lead to the build-up of elliptical galaxies
with hot X-ray halos which will later be incorporated into clusters \citep[e.g.][]{PBnat,MZ99,SNDS}.
Groups are also expected to be important because they are the first
level of the super-galactic hierarchy.  Models of galaxy formation assume the rate at which gas falls
onto a galaxy depends on its environment \citep[e.g.][]{Cole2000,HS03}.  In these models,
satellite galaxies do not have their own supply of hot gas to replenish
gas in the disk used to form stars.  Thus, star formation will
gradually decline in any galaxy which is part of
a larger halo \citep{infall,Diaferio}; groups will be the first
environment to demonstrate this effect.

\subsection{Previous Work}
Our understanding of galaxies in groups and lower density environments
has been hindered by the difficulties in obtaining large, unbiased samples.
Most of the work has been restricted to compact groups
\citep[e.g.][]{Hickson,RHF,IPV,Coziol,N+00,VM01,dlR01,KF03}, which are most likely to be physically bound
systems.  However, these may be a special class of group, not
representative of a typical stage of the hierarchy through which
most galaxies evolve.

\citet{PG84} analysed the morphology-density relation in groups
selected from the CfA redshift survey \citep{Cfa3} and demonstrated that the original relation found
in clusters by \citet{Dressler} extends continuously to lower
densities.  Below a density of $\sim 1$ Mpc$^{-3}$ ($M_{B(0)}<-17.5$) no
further correlation is seen.  Similar trends have since been confirmed
in numerous other studies \citep[e.g.][]{ZM98,HO99,Tran,Treu03} .  Interestingly, \citet{PG84}
found the trend with local density is the same in both rich clusters and poor
groups; the only difference is that poor groups typically sample lower
density regions and, thus, have populations more dominated by late-type
galaxies.  It has been argued based on this evidence that galaxy-galaxy interactions
within the group environment are the mechanism responsible for the high
fractions of early-type galaxies in clusters \citep[e.g.][]{ZM98,ZM00,HO00}.

Less work has been done on the stellar populations of group galaxies, though 
there is good evidence that they are intermediate between those of the field and
rich clusters \citep{A+93,H+98,G+02,Tran,CNOC_groups2_short}.
More recently, analysis of SFRs in galaxy groups
selected from  the partially-completed 2dFGRS survey
has been done by \citet{MZ2dF}.  Based on this group catalogue,
\citet{Martinez} found that even the lowest-mass groups, with masses
$\sim10^{13}M_\odot$, show reduced total SFRs relative to
the field.   \citet{D+02} claim that the spectral-type dependence on local density
is only observed in groups more massive than $3\times10^{13}M_\odot$;
however, this claim appears to arise mostly from the fact that their
low-mass groups do not sample densities as high as found in the
higher-mass groups. 

\subsection{The purpose of this paper}
We analyse the local correlation between star-formation activity and
environment using data obtained from the two largest
galaxy surveys ever conducted: the 2dFGRS
\citep{2dF_colless,2dF_final} and the SDSS \citep{SDSS_tech_short}.  In
particular, we have selected
a sample of galaxy groups from each survey, but in significantly
different ways.  Since the definition of a group is partly
subjective, and a variety of algorithms have been developed to find
groups in redshift surveys,
comparing results from the two catalogues allows us to 
investigate the sensitivity of our results to the way the group
catalogue is constructed.  

The purpose of this paper is to establish how star formation in the
galaxy population, as characterised by the distribution of H$\alpha$
emission, depends on environment.  Specifically, we will investigate 
whether the most important variable is 1) velocity dispersion of the
embedding group or cluster; 2) local galaxy density, on scales
$\lesssim 1$ Mpc; or 3) large-scale structure, as parameterised by the
density on $\sim 5$ Mpc scales.  We will show that the number of
actively star-forming galaxies depends on both the local and
large-scale density, but that the properties of the star-forming
galaxies themselves do not.  

A summary of the paper follows.  In Section~\ref{sec-data} we describe the
details of the two datasets analysed herein, focusing on the
construction of group catalogues, measurement of emission line
strengths, and sample selection.  
Our results are presented in Section~\ref{sec-results}.  We discuss the physical
implications of our results in Section~\ref{sec-discuss}, and draw some
conclusions in Section~\ref{sec-conc}.
Throughout the paper we use a cosmology of $\Omega_m=0.3$,
$\Omega_\Lambda=0.7$ and $H_0=100h\kmsmpc$, with
$h=0.7$.  All distances are proper lengths in units of Mpc.

\section{Data}\label{sec-data}
We will use local data gleaned from the two largest redshift surveys
available: the 2dFGRS  \citep{2dF_colless,2dF_final} and the SDSS \citep{SDSS_tech_short}.  Below we summarise the
data in each survey, and describe our group catalogues, emission line
measurements, and sample selection criteria.

\subsection{The 2dF Galaxy Redshift Survey}
\subsubsection{Summary of the data}
The 2dFGRS  has obtained over 220\,000 spectra of 
galaxies selected in the photographic $b_{\rm J}$ band, from the APM galaxy
catalogue.  The targeted galaxies are located in two contiguous declination strips, plus 99
randomly located fields. One strip is in the southern Galactic 
hemisphere and covers approximately 80${^\circ}\times15{^\circ}$ centred 
close to the SGP. The other strip is in the northern Galactic hemisphere 
and covers 75${^\circ}\times10{^\circ}$. The 99 random fields are located 
over the entire region of the APM galaxy catalogue in the southern 
Galactic hemisphere outside of the main survey strip. We only use the
contiguous fields for this work.  Full details of the 
survey strategy are given in \citet{2dF_colless}.

The survey spectra were obtained through $\sim 2$\arcsec\ fibres, and 
cover the wavelength range 3600--8000\AA\ at 9\AA\ 
resolution. Only the wavelength range of 3600--7700\AA\ is used during the 
line fitting procedure due to poor signal to noise and strong sky emission 
in the red part of the spectrum. The wide wavelength range is made 
possible by the use of an atmospheric dispersion compensator (ADC) within 
the 2dF instrument \citep{2dF_short}.  The accuracy of each
individual redshift is $\sim 85\kms$ \citep{2dF_colless}.

\subsubsection{The group catalogue}\label{sec-2dfgroups}
The group catalogue is based on a friends-of-friends percolation
algorithm, which
links neighbouring galaxies together if they lie within a
specified linking-length of each other \citep{Eke-groups}.  The linking-length is scaled
with redshift in order to obtain groups of a constant overdensity in the
magnitude limited survey.  This is done by scaling the length according
to $n(z)^{-1/3}$, where $n(z)$ is the mean galaxy density at redshift
$z$.  Furthermore, the linking length along the line of sight ($\ell_{||}$)
is allowed to be larger than that in the plane of the sky ($\ell_\perp$), to account for the effects of
peculiar velocities.  The algorithm is extensively tested on mock
catalogues derived from numerical simulations with $\sigma_8=0.9$, so the completeness and
contamination of the catalogue is understood.  For details we refer the
reader to \citep{Eke-groups}; we briefly review the salient points
here.  The mock galaxy catalogues are created using the semi-analytic
models of \citet{Cole2000}, with the 2dFGRS  selection function.  Groups
identified from this mock catalogue are then compared with the
corresponding ``true'' group, identified as galaxies populating dark
matter halos identified using a friends-of-friends algorithm with a
linking length of $b=0.2$ times the mean particle separation.  The
parameters of the
group-finding algorithm are tuned to provide the best match between the
median properties (size and mass) of the observed and true
groups.  These best-fit parameters are $b=\ell_\perp n(z)^{1/3}=0.13$, $L_{\perp,
  \rm max}=2h^{-1}$ Mpc and $R=11$, where $L_{\perp,  \rm max}$ is the
maximum linking-length permitted across the line of sight and
$R=\ell_{||}/\ell_\perp$.  These values are insensitive to
$\sigma_8$ for reasonable values of the normalisation.  Small, parameterised perturbations of $b$ and $R$
are allowed to remove small differences in the recovered properties
that correlate with halo mass. 
The median recovered velocity dispersions are accurate to better than $\sim 10$\%,
independent of halo mass, when compared with the velocity dispersion of
the parent dark matter halo; however, the scatter in this accuracy is
large.  The catalogue is highly complete, recovering $>95$\%
of halos with dark matter mass $M\lesssim4\times10^{14}$ $M_\odot$; the price
to be paid is some contamination from unphysical systems.  
For the most massive haloes, the algorithm is susceptible to a small
amount of fragmentation; approximately
10\% of haloes
with $M\gtrsim10^{14} M_\odot$ are fragmented into more than one group
with mass at least 20 per cent that of the parent dark matter halo.

For each group, we calculate the velocity dispersion using the gapper estimate of
\citet{WT76}, as discussed in \citet{Beers}, which is insensitive to
outliers.  The group centre is computed by iteratively rejecting the
most distant galaxy until only two galaxies remain; the centre is taken
to be the position of the brighter of these two galaxies.

\subsubsection{H$\alpha$ as a star formation tracer in the 2dFGRS}\label{sec-2dfsfr}
The SFR is directly related to the H$\alpha$ emission luminosity
\citep[e.g.][]{K83}, and we will use this emission line as our tracer of star
formation.  To provide a reliable description of SFR, however, H$\alpha$
luminosities need to be corrected for underlying absorption, dust
extinction, and aperture effects, all of which are important
\citep[e.g.][]{CL,Hopkins01,Hopkins03,Afonso}.  When these corrections are
made, there is good agreement between H$\alpha$--derived SFRs, and
those derived from the far infrared, radio or ultra--violet continuum
\citep{Hopkins03}.  In this paper, however, we will focus only on the
rest-frame equivalent width of the H$\alpha$ line, \ewha, corrected
for underlying absorption (see below).  Uniform dust extinction
will not affect the \ewha, though selective extinction around massive
stars will \citep{CL}.  Aperture corrections to the {\it flux} can be
substantial, typically ranging from a factor $\sim2$ to $\sim6$
\citep{Hopkins03}; however, the effect of this missing flux on \ewha\
depends on the spatial distribution of the H$\alpha$ emission.
As long as the size distribution of galaxies is independent of
environment, neglecting this correction does not affect the results of
this paper.  We will demonstrate that this assumption is justified, in
Appendix~\ref{sec-apeffects}.

As described in Paper~I, all of the measurements of line equivalent widths have been performed using a
completely automatic procedure.   In summary, up to 20 individual absorption and emission lines are fitted
simultaneously with Gaussian profiles.  The H$\alpha$ emission line is accurately deblended from the adjacent
[N{\sc ii}]$\lambda6548$\AA\ and [N{\sc ii}]$\lambda$6583\AA\ lines; the [N{\sc ii}] lines are
constrained to be in emission while the H$\alpha$ line may be either
emission or absorption (but not both); measurements are made whether or
not the line is detected in emission.  We will add 1\AA\ to the H$\alpha$ equivalent widths to
approximately account for the effects of stellar absorption \citep{Hopkins03}.  This correction is not important for
galaxies with significant emission, which are of interest here,
but ensures that the mean \ewha\ is never much less than zero.
Because of the uncertain flux calibration of the 2dFGRS  spectra, we do not derive H$\alpha$ luminosities,
or SFRs, but restrict the analysis to the observable quantity \ewha.  

We also wish to exclude from the analysis galaxies in which most of the
emission comes from an active galactic nuclei (AGN).  Since the flux
calibration is not reliable over long wavelengths, we just use the
criterion  [N{\sc ii}]$\lambda6583$/H$\alpha>0.63$ to identify
AGN-dominated spectra, when the equivalent widths of both lines are
greater than 2 \AA\ \citep[e.g.][]{Miller_agn}.  Since the AGN fraction
does not appear to correlate with environment, however
\citep{Miller_agn}, the correction does not affect any of the
conclusions in this paper.

\subsubsection{Sample selection}
We select a volume-limited subset of the contiguous fields, where the lower redshift limit, $z>0.05$, is
chosen to reduce our sensitivity to aperture effects.  
The upper
redshift limit, $z<0.095$, is chosen because measurement 
of H$\alpha$ emission becomes difficult when it is redshifted beyond
$\sim 7200$ \AA\ and night sky emission lines are strong\footnote{In
  Paper~I we used an upper limit of $z=0.1$; the small
  change here is made for consistency with the SDSS sample as presented
in Paper~II and this paper.}.
At this redshift, the magnitude limit of the survey ($b_{\rm J}=19.45$)
corresponds to a k-corrected luminosity of $M_b\approx -19$.  However,
for consistency with the SDSS sample (see Appendix~\ref{sec-cfapp}), we
limit this sample to $M_b=-19.5$, using the average k-correction of
\citet{2dF-lf2_short}.  Finally, we exclude galaxies that are within
$2$ Mpc of a survey boundary; more stringent cuts are made when
computing densities on various scales, as appropriate.

\begin{figure}
\leavevmode \epsfysize=8cm \epsfbox{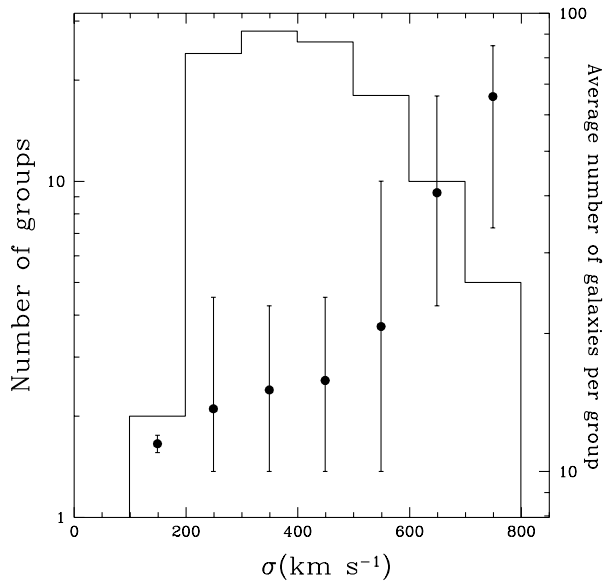}
\caption{The histogram shows the abundance of groups with at least ten
  members in the 2dFGRS 
  sample as a function
  of velocity dispersion, corresponding to the left axis.  The {\it
    solid points} are the average number of galaxies (brighter than
  $M_b=-19$) per group, with values corresponding to the right axis.
  The error bars span the range from the 10th to 90th percentile of the
  distribution.
\label{fig-2dFgroups}}
\end{figure}

Our final,  volume-limited sample 
contains 20154 galaxies.  
For computations of \ewha, we exclude galaxies in which the continuum was negative,
or a Gaussian was a poor fit to the line (see Paper~I).  Furthermore, we
restrict our analysis to data taken after August 1999, since the
extreme ends of spectra obtained earlier are severely affected by 
problems with the ADC \citep{2dF_short}.  These restrictions reduce the
usable sample to
12683 galaxies.  Of these, 846 (6.7\%) are identified as AGN (see
Section~\ref{sec-2dfsfr}) and excluded from the sample.
This leaves 11837 galaxies, of which 7012 (59 per cent) are in the
friends-of-friends group catalogue; 1577 (13 per cent) are in groups
with at least ten members above the luminosity limit, which we use for
our analysis.  This latter requirement is made so that meaningful
estimates of local density can be made.
We use the friends-of-friends linking algorithm to designate group
membership, rather than selecting all galaxies within some specified
radius of a chosen centre.  No further selection of the clusters is
made, and some of the systems are therefore dynamically unrelaxed
groups, for which the velocity distribution is not a reliable indicator
of dynamical mass.
The velocity dispersion distribution of groups with more than ten bright members is shown
in Fig.~\ref{fig-2dFgroups}.  
As described in Section~\ref{sec-2dfgroups}, this is designed to be a very highly
complete catalogue that will still have some contamination from
unphysical systems, though this contamination is substantially reduced by our selection of
groups with at least ten members \citep{Eke-groups}.
\subsection{The first SDSS data release}
\subsubsection{Summary of the data}
The SDSS (http://www.sdss.org) is a joint, 5 passband
($u$, $g$, $r$, $i$, $z$), imaging and medium--resolution
($R\simeq1800$) spectroscopic survey of the Northern Galactic Hemisphere
\citep[see][for details]{SDSS_tech_short}. In May 2003, the SDSS publicly
released the first official set of data, named DR1, which comprises 186,240 spectra
of galaxies, stars and QSOs over 1360 ${\rm deg^2}$ of sky. This release is
fully described in \citet{DR1}, and
the reader is referred to \citet{Strauss02_short} for a detailed description of
the spectroscopic target selection for the SDSS main galaxy survey. 

The spectra are obtained from 3\arcsec\ diameter fibres, larger than
those of the 2dFGRS.  The spectrographs produce data covering 
3800-9200 \AA, with the beam split at 6150 \AA\ by a
dichroic. The spectral resolution at $\lambda\sim 5000$\AA\ is $\sim
2.5$\AA, and redshift uncertainties are $\sim 30\kms$.

\subsubsection{The group catalogue}
Groups are selected from the SDSS in a fundamentally different way from
2dFGRS  groups (cf. Section~\ref{sec-2dfgroups}), which allows us to test the
sensitivity of our results to the group-finding algorithm.
The SDSS algorithm (Nichol, Miller et al. in prep) is a semi-parametric,
high dimensional technique developed to find galaxies clustered in
both position and colour. The premise is that galaxies within clusters
evolved similarly, and thus galaxy clusters contain subsets of galaxies
that have similar spectral energy distributions (SED). Note that this
is not a red-sequence finder \citep[e.g.][]{GY00}, and will detect
groups of blue galaxies if they have similar colours.  Galaxy colours
are used as a proxy for the shape of the SED since they cover a larger
wavelength range than the spectra and because the dimensionality of the
problem is reduced to a manageable 7-d space (two spatial positions, redshift,
$u-g$, $g-r$, $r-i$, and $i-z$). In practice, we expect clustering in colour-space
to be a signal from the red sequence \citep[e.g.][]{GY00}. 
Galaxy overdensities in this 7-d space are found
by comparing to random locations in the SDSS DR1 survey.
By design, the algorithm is extremely pure (i.e. few false positives).
This high purity is the result of  the lack of projection in 7-d space,
and the use of the False Discovery Rate statistical method in choosing
a threshold above which galaxies are considered clustered \citep[see][]{Miller01}.

The algorithm has been tested against mock catalogues 
(Wechsler et al., in prep), in which galaxies are assigned 
$r$-band magnitudes and placed onto dark matter particles in such a way
that the luminosity-dependent clustering of the SDSS seen in
\citet{Blanton03} is matched. The remaining magnitudes  are
then added in so that the correlation between colour and local density, as
observed in the SDSS,
is also matched. The same cluster-finding algorithm used on
the real data is then run on the mock catalogues. Preliminary tests show that
the cluster catalogue is $>90$\% complete for clusters with
mass $M>2.5\times 10^{14} M_\odot$; the completeness decreases toward
lower masses. 
The catalogue is 100\% pure for systems with $M>5\times 10^{14} M_\odot$ and always greater
 than 90\% pure (Nichol, Miller et al., in prep.). 
We exclude groups in which the velocity distribution (measured using
all galaxies, not only those identified as clustered in colour space) is significantly
different from a Gaussian, so that we preferentially select dynamically
isolated, relaxed systems.  The final catalogue contains
104 bona-fide clusters with reliable velocity dispersions that are
likely indicative of system mass.  This
catalogue is a good complement to the 2dFGRS  catalogue, as it is a
highly pure catalogue (i.e., with little contamination), at the expense
of being incomplete for low-mass groups.  All galaxies within
$1000\kms$ and twice the virial radius (defined by the galaxy number
overdensity relative to the field) of the cluster are considered
group members; the exact radius chosen is not important because we
present our data as a function of local density, which correlates well
with radius within the virialised region.

\subsubsection{H$\alpha$ as a star formation tracer in the SDSS}
We will again use the H$\alpha$ equivalent width as
a tracer of star formation, as we have done 
for the 2dFGRS  sample (see Section~\ref{sec-2dfsfr}).
Emission and absorption lines are measured automatically from the
spectra by fitting multiple Gaussians where required.  Again a measurement is
provided for every line, whether or not it was detected in emission, and
we make a 1\AA\
correction for underlying stellar absorption \citep{Hopkins03}.  
The effect of converting \ewha\ to SFR using various conversions
\citep[e.g.,][]{Kenn_review,Hopkins01,CL}, were discussed in
Paper~II and will not be considered here.

For the SDSS spectra we can more effectively exclude AGN (relative to
the 2dFGRS), since the spectra are accurately flux calibrated.  Here, we will consider the [N{\sc
  ii}]$\lambda6583$/H$\alpha$ and [O{\sc iii}]/H$\beta$ ratios, using
the classification of \citet{Miller_agn}.  Where
possible, all four lines are used to identify AGN; if only one of the
ratios is available, then only that one is used.  Most (87\%)  of the
AGN are identified
based on the [N{\sc ii}]/H$\alpha$ ratio alone.

\subsubsection{Sample selection}\label{sec-sdsssample}
We will use the same volume-limited sample as in
Paper~II.  The main criteria for selection are 
$0.05<z<0.095$ and $M_r<-20.6$ (k--corrected, for ${\rm H_0} = 70\kmsmpc$).  
As for the 2dFGRS  data, the lower redshift
limit is imposed to minimise aperture bias due to large
nearby galaxies (see Appendix~\ref{sec-apeffects} for more detail).   
The upper redshift limit is that where  our luminosity
limit equals the magnitude limit of the SDSS
\citep[$r=17.7$;][]{Strauss02_short}.
From a volume--limited sample of 19287 galaxies we exclude 4237 that
are within $2$ Mpc of a survey boundary, and another 64 galaxies for
which no reliable measurement of H$\alpha$ is available, leaving 14986
galaxies.  
Finally, removing a $\sim 12\%$ AGN contribution we have 13131 galaxies in the final sample.
Of these, 1939 (15\%) are associated with a group in the catalogue.

\begin{figure}
\leavevmode \epsfysize=8cm \epsfbox{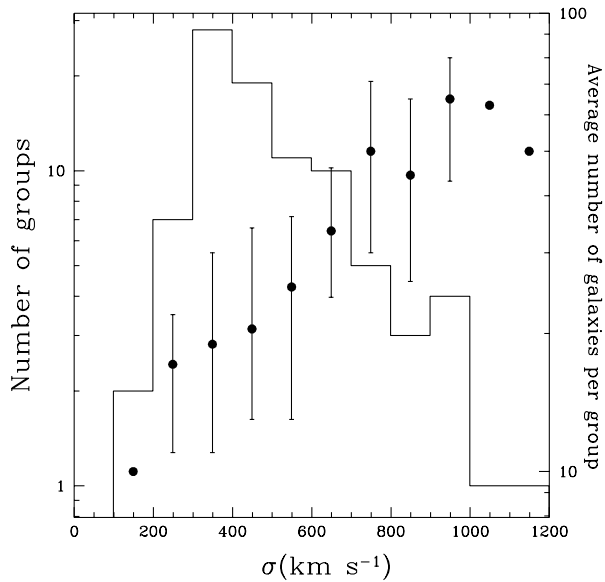}
\caption{As Fig.~\ref{fig-2dFgroups}, but for groups in the SDSS
  sample, restricted to groups with at least ten member galaxies brighter than $M_r=-20.6$.
\label{fig-SDSSgroups}}
\end{figure}

In Fig.~\ref{fig-SDSSgroups}   we show the
velocity dispersion distribution for the selected subset of SDSS groups.
The distribution is different from that of the
2dFGRS groups (Fig.~\ref{fig-2dFgroups}), likely due to the fact that the former catalogue is incomplete
at the lowest velocity dispersions, while the latter has more
contamination.  Furthermore, the SDSS catalogue includes a few clusters
with very high velocity dispersions, $\sigma>800\kms$, while such large
systems tend to be fragmented in the 2dFGRS catalogue.
\subsection{Homogeneity of the samples}
In using the 2dFGRS  and SDSS data together, we need to ensure that the
samples are comparable in terms of selection and derived measurements.
This detailed comparison is made  in Appendix~\ref{sec-cfapp}.
The only difference relevant to our results is that the $b_{\rm J}$-selection of the
2dFGRS  results in a sample that is more biased toward galaxies with large \ewha\
than the $r-$selected SDSS.  Because of this difference, we will show
most of our results for the two surveys separately.
The final combined sample contains 24968 galaxies and 204 groups with
at least ten members.  Approximately 14\% of all galaxies in the sample
are associated with a group.

\subsection{Density Estimates}\label{sec-densities}
The local galaxy number density around a given galaxy is not in itself a
well-defined quantity.  As can be anticipated from the shape of the
correlation function \citep[e.g.][]{Baugh99}, the
density of neighbouring galaxies tends to increase as you probe closer
to the target galaxy (see Appendix~\ref{sec-kde}).    
We are therefore faced
with two choices when characterising the density around a galaxy.  One
is to measure the density within
a fixed distance scale.  This ensures that both high- and low-density
regions are measured at the same scale; however, the measurement is noisy and of limited
dynamic range due to the arbitrarily small, and finite, number of
galaxies within the chosen distance.  On the other hand,
we can use a  systematically larger
scale in lower density regions.  This improves sensitivity and
precision at low densities, but may be difficult to interpret because
high- and low-density environments are measured at different physical scales.
Without prior knowledge of the nature of the relevant physical effects,
we cannot say which is the more meaningful estimate.  For example, if
we believe galaxy-galaxy interactions are important, then perhaps it is
only the distance to the nearest neighbour that is relevant \citep{Lambas}.  It is easy to imagine scenarios
where either the distance to the N$^{th}$ neighbour, or density
measured on a physical scale, is likely to be more meaningful.  

In this paper, we will adopt two density estimators, which are
described in Appendix~\ref{sec-app2}.  The first is a
traditional projected density estimate, $\Sigma_5$, which is based on
the projected distance to the fifth-nearest neighbour within $\pm 1000\kms$.  
Inasmuch as we are willing to believe that galaxies can
be found in discrete, relatively isolated groups, we must be cautious
about interpreting our density estimate in groups where the number of
members is less five.  
For a group with only four members, the fifth nearest neighbour will
clearly not lie in the same group and, therefore, the density computed
from this distance may not be what is wanted.  To ensure that we
maintain our intuitive idea of a local environment within groups,
we only consider groups which have at least ten galaxies brighter
than our luminosity limit.  Furthermore, we exclude any galaxies in
which the fifth--nearest neighbour is more distant than the closest
survey boundary, or within $1000\kms$ of our redshift limits, to ensure
accurate measurements.  This may bias us against finding 
low-density regions, but does not affect the observed trends of galaxy
property with $\Sigma_5$ \citep[e.g.][]{Miller_agn}.  

The second estimator is a three-dimensional density $\rho_\theta$, 
obtained by convolving the galaxy field with a Gaussian of dispersion
$\theta$.   In particular, we will consider the density measured on scales
$\theta=1.1$ and $5.5$ Mpc (see Appendix~\ref{sec-app2}).  
These measurements underestimate the density in clusters with
large velocity dispersions, but are particularly useful for probing low
density regimes where peculiar velocities are small.  The main
disadvantage is that the signal-to-noise ratio varies with density, and
is low at low densities, when there are few galaxies within the
aperture.  In our analysis we will therefore indicate the density at
which there is less than one galaxy within the Gaussian dispersion
$\theta$, and will exclude any galaxies which are located less than
$2\theta$ from a survey boundary, to avoid biassing the density estimate.

\section{Results}\label{sec-results}
\subsection{Dependence of \ewha\ on environment}\label{sec-res-sfrden}
In Papers~I and~II we showed that there is a strong correlation
between \ewha\ distribution and local galaxy density.  In this section
we will first define precisely what property of the galaxy distribution
correlates with environment; then proceed to explore how this correlation
depends on the definition of local environment.
  
\subsubsection{Bimodality in the \ewha\ distribution}
\begin{figure}
\leavevmode \epsfysize=8cm \epsfbox{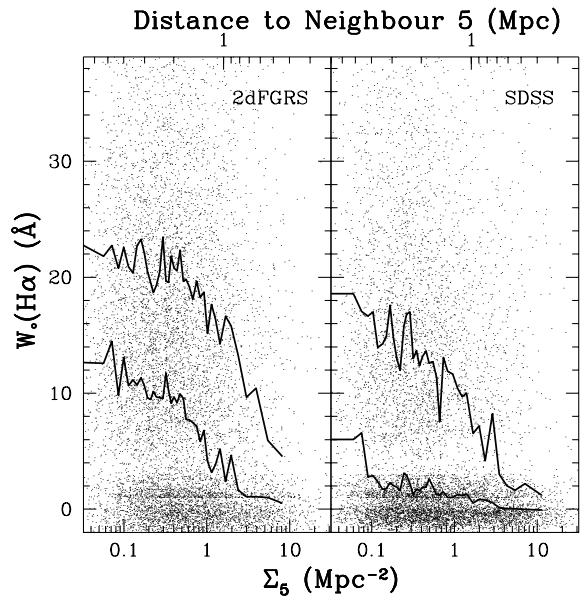} 
\caption{The dependence of \ewha\ on projected, local density for the 2dFGRS  {\it
    (left panel)} and SDSS {\it (right panel)} samples.  The top axis
  shows the distance to the 5$^{th}$ nearest neighbour, from which
  $\Sigma_5$ is computed.  The {\it
    solid lines} show the median and \75th\ percentile of \ewha, in bins of
  100 galaxies.  Although we only plot the data for \ewha$<40$\AA, they
  extend to \ewha$\sim100$\AA; see Fig.~\ref{fig-hadist} for the full \ewha\
  distributions as histograms.
\label{fig-sfrden1}}
\end{figure}

In Fig.~\ref{fig-sfrden1} we present the correlation between \ewha\ and
$\Sigma_5$ for the 2dFGRS  and SDSS samples. On the top axis we show
the distance to the fifth--nearest neighbour corresponding to
$\Sigma_5$, which shows that at the lowest densities we are measuring the galaxy
distribution on $\gtrsim3$ Mpc scales, while in the densest regions the
measurement is made at $\lesssim 300$ kpc.  Recall (Section~\ref{sec-densities})
that we exclude any galaxy for
which the fifth--nearest neighbour is more distant than the nearest
survey boundary, or within $1000\kms$ of our redshift limits.  The solid lines
show the median and \75th\ percentile of the \ewha\ distribution.  In
both cases, we reproduce the results of Paper~I and Paper~II; the
distribution changes distinctly, at a density $\Sigma_5\sim 2$
Mpc$^{-2}$.  This is characterised by the near-total lack of galaxies with large \ewha\
at densities greater than this value.  At lower densities, there
remains a correlation between \ewha\ and $\Sigma_5$; however it is
weak, with the average \ewha\ increasing by only $\sim 25$ per cent
over an order of magnitude in local density.  The difference in normalisation between the SDSS and
2dFGRS  surveys is due to the  different selection criteria for the
spectroscopic sample,
as discussed in Appendix~\ref{sec-cfapp}. 

The \ewha\ distribution shown in Fig.~\ref{fig-sfrden1} reveals the
presence of two distinct galaxy populations: those with significant,
ongoing star formation covering a broad range in H$\alpha$ strength
(from $\sim 4$\AA\ to $>40$\AA), and those with no ongoing star
formation, which form a conspicuous horizontal ridge line about
\ewha$=0$.  (This is more clearly seen in Fig.~\ref{fig-hadist}, where we
show the \ewha\ distributions as histograms).
This recalls the bimodality observed in the
colour distribution \citep{Strateva01_short,Blanton03BB};
\citet{Baldry03} find that the red peak distribution can be explained as the result
of mergers between galaxies in the blue peak.
We will therefore focus our analysis on the
star-forming population, and its variation with environment. 

\begin{figure}
\leavevmode \epsfysize=8cm \epsfbox{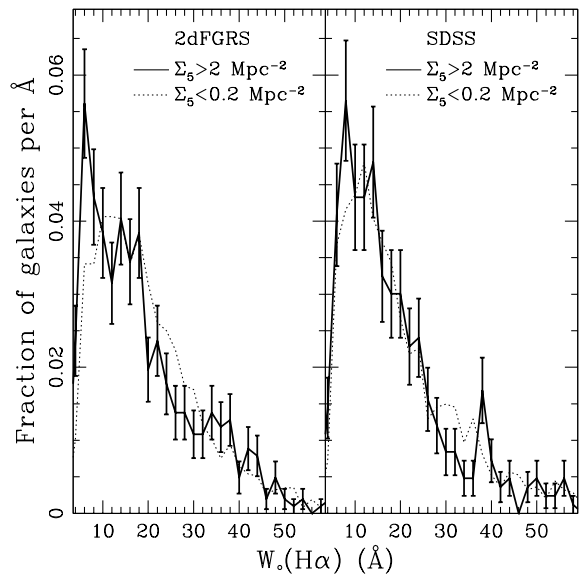}
\caption{The \ewha\ distribution for galaxies in the 2dFGRS {\it (left
    panel)} and SDSS {\it (right panel)} with
  \ewha$>4$\AA, in low-density environments
  ($\Sigma_5<0.2$ Mpc$^{-2}$, {\it dotted line}) and high-density
  environments ($\Sigma_5>2$ Mpc$^{-2}$, {\it solid line}).  We show
  Poisson-distributed error bars on the high-density subsample, which
  is the smaller of the two.  
\label{fig-half}}
\end{figure}

In
Fig.~\ref{fig-half} we show the \ewha\ distribution only for those
galaxies with \ewha$>4$ \AA, in environments  with the highest and
lowest $\Sigma_5$ densities.
The distributions are very similar; a
Kolmogorov--Smirnov test cannot reject the hypothesis that the low- and
high-density distributions are drawn from the same population with more than
1$\sigma$ confidence.   Any difference, no matter how significant, must
be small: the mean \ewha\ is $\sim 21.5$\AA\ and $\sim 20$\AA\ for the low- and high-density
populations, respectively, in both the 2dFGRS and SDSS samples.
Thus, the observed
trend of mean \ewha\ with density (Fig.~\ref{fig-sfrden1}) is due almost  
entirely to the relative proportion of galaxies with \ewha$>4$ \AA; 
there is at most weak sensitivity to
environment within the actively star forming population alone\footnote{We will not deal
directly with the distribution of \ewha\ in galaxies with \ewha$<4$
\AA, since this distribution is most likely dominated by measurement
uncertainties, including systematic effects like stellar absorption,
rather than star formation activity.}.

\subsubsection{The abundance of star-forming galaxies}
\begin{figure}
\leavevmode \epsfysize=8cm \epsfbox{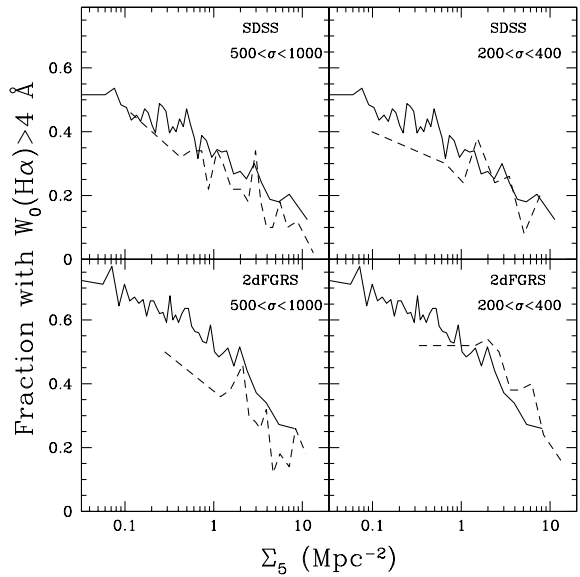}
\caption{The fraction of galaxies with \ewha$>4$ \AA\ as a function of local density for the 2dFGRS  {\it
    (bottom panels)} and SDSS {\it (top panels)} samples.  The {\it
    solid lines} represent the full
galaxy sample, in bins each containing 250 galaxies.  The {\it dashed
  lines} are restricted to galaxies which lie in groups or clusters
with the indicated velocity dispersion, in bins of 50 galaxies each.
Poisson-distributed uncertainties are typically $\sim 0.1$ on the {\it dashed lines}
and $\sim 0.05$ on the {\it solid lines}.
\label{fig-haden}}
\end{figure}

Motivated by the results of the previous subsection, we show,
in Fig.~\ref{fig-haden}, how the fraction of galaxies with
\ewha$>4$ \AA\ depends 
on $\Sigma_5$.  It is evident that the fraction decreases steadily with
increasing density; the break at $\Sigma_5\sim2$ Mpc$^{-2}$ is
still there, but less apparent than in Fig.~\ref{fig-sfrden1}.  

Since the identification of a characteristic density can have important
implications, it is useful to understand why it
appears so strongly in Fig.~\ref{fig-sfrden1} (and Papers~I and~II) but
is much weaker in Fig.~\ref{fig-haden}.  
Upon close examination,
a change in slope is most evident in the median of the 2dFGRS, at $\Sigma_5\sim
1$ Mpc$^{-2}$, and in the \75th\ percentile of the SDSS, at
$\Sigma_5\sim 3$ Mpc$^{-2}$.  From Fig.~\ref{fig-haden} we see
that these two densities correspond to the point at which
the fraction of galaxies with \ewha$>4$\AA\ drops below 50
(2dFGRS) or 25 (SDSS) per cent; at higher densities the median and
\75th\ percentile, respectively, are tracing the non-star forming
population with \ewha$<4$\AA, and the trend with $\Sigma_5$ largely disappears.

It is worth stressing that the trend with $\Sigma_5$ is fairly
insensitive to projection effects, despite the fact that we are
projecting over a $\pm 14$ Mpc cylinder, and despite the nonuniform
density of the background, as we demonstrate in
Appendix~\ref{sec-proj}.  To summarize the results of that Appendix, at $\Sigma_5\sim 0.1$
Mpc$^{-2}$, where the projected contamination is $\sim 50$ per cent, we
only overestimate the emission-line fraction by about 5
per cent.  This is because the emission-line fraction varies weakly
with $\Sigma_5$ at low densities, so there is little physical
difference between the properties of the target galaxies and the
projected galaxies.  At high $\Sigma_5$, the projected fraction drops
strongly, both because the contrast with the field is increasing, and
because the radius of the projected cylinder is decreasing.  
Thus, the observed
emission-line fraction remains within $\sim 5$ per cent of the true
value, despite the fact that the contrast between the target and
projected galaxy populations becomes large.  The observed trend with
$\Sigma_5$ is, therefore, a physical one.

Fig.~\ref{fig-haden} also shows the correlation for galaxies in groups
of different velocity dispersion.  We note that, although
velocity dispersion gives a good indication of the group mass on
average, there is a significant scatter arising because our groups
are not all relaxed, spherical systems.  Therefore, the velocity
dispersion in some cases will be a better indicator of substructure or dynamic
state than of mass.  
The correlation of emission-line fraction with
density is present in both high- and low-velocity dispersion groups,
selected from either the SDSS or 2dFGRS.  
There is marginal evidence (particularly in the 2dFGRS sample) that galaxies
in the highest-velocity dispersion clusters have 
a low fraction  of galaxies with \ewha$>4$\AA\  for their local density, relative to galaxies in
lower-dispersion groups.  Projection will have a more complex effect on
this result, since the physical size of the group depends on $\sigma$;
in particular, we expect the overestimation of 
both $\Sigma_5$ and \ewha\ to be greater in
high-$\sigma$ clusters than in low-$\sigma$ clusters.  This means the
high-$\sigma$ cluster \ewha\ distribution should exceed
that of low-$\sigma$ groups at a fixed $\Sigma_5$; this is opposite to what we
find, and cannot therefore be the explanation for the small observed
difference.  Finally we note the remarkable similarity between the
results for $\sigma<400\kms$ groups in both samples.  Since the 2dFGRS
catalogue is highly complete, and the SDSS catalogue is highly pure,
the persistence of a strong \ewha--$\Sigma_5$ relation in groups from
both catalogues is not likely due to a selection effect.
\begin{figure}
\leavevmode \epsfysize=8cm \epsfbox{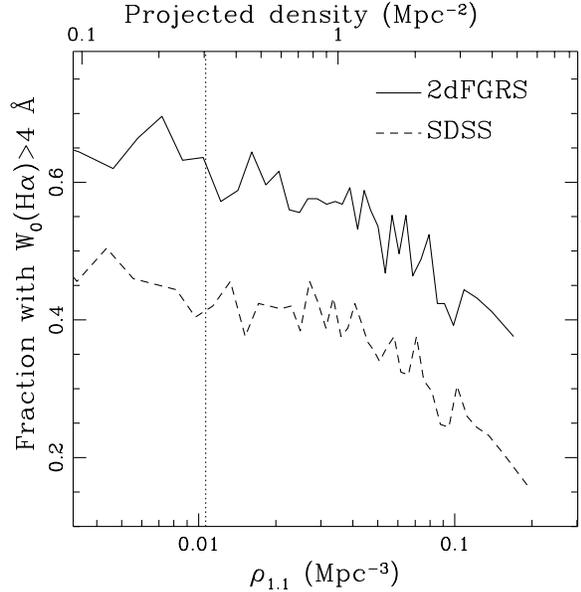}  
\caption{The fraction of galaxies with \ewha$>4$\AA\ in the SDSS {\it
    (dashed line)} and 2dFGRS {\it (solid line)},  as a function of three-dimensional
  density $\rho_{1.1}$, estimated with a Gaussian kernel with 1.1 Mpc
  standard deviation.  Each bin contains 250 galaxies.  The vertical, {\it dotted line}
corresponds to the density where no galaxy lies within the 
Gaussian filtering scale $\theta$, making the density estimate
particularly noisy.
On the top axis
we show the estimated two-dimensional projected density within a
cylinder of length $2000\kms$, for comparison with $\Sigma_5$.
\label{fig-kde}}
\end{figure}

We now consider the
three dimensional Gaussian kernel density estimator with a 1.1 Mpc
standard deviation, $\rho_{1.1}$ (the motivation for the choice of filtering
size is given in Appendix~\ref{sec-app2}).  
The $\rho_{1.1}$ density measure  allows us to find
galaxies in the very lowest density environments, which are not easily
measured from $\Sigma_5$ since that quantity projects the galaxy distribution
over a $2000\kms$ cylinder\footnote{This projection corresponds
to $\sim 25$ Mpc, much larger than the linear size of
typical void regions \citep{Benson_voids}.}.   In our analysis, we
exclude galaxies that are within
$2.2$ Mpc of a survey boundary, for which $\rho_{1.1}$ cannot be
reliably determined; we also note that $\rho_{1.1}$ becomes noisy
when there are no galaxies within the filtering scale 1.1 Mpc,
$\rho_{1.1}<0.01$ Mpc$^{-3}$.  

Fig.~\ref{fig-kde}
shows how the fraction of galaxies with \ewha$>4$\AA\ depends on
$\rho_{1.1}$.  To approximately compare this with the corresponding
$\Sigma_5$ value,  we compute the
projected density within a cylinder of length 2$\times1000\kms$ along
the line of sight, to obtain an equivalent surface density, shown along
the top axis.  The emission-line fraction shows a
strong dependence on $\rho_{1.1}$, and gets significantly
steeper at $\rho_{1.1}\gtrsim0.05$ Mpc$^{-3}$.   However, the
correlation is still 
present at the lowest densities observed.  It is not clear whether the
change in slope at $\rho_{1.1}\gtrsim0.05$ Mpc$^{-3}$ highlights an
interesting physical scale, or if $\rho_{1.1}$ simply loses sensitivity
to the underlying density distribution at low densities, an effect that
would be exaggerated by the logarithmic scale.  Furthermore, the highest
densities (which tend to correspond to clusters with high velocity
dispersions) will generally be underestimated as a result of large
peculiar velocities, and this may be partly
responsible for the change in slope.

\subsection{Local or Global Density?}\label{sec-lgd}
\begin{figure}
\leavevmode \epsfysize=8cm \epsfbox{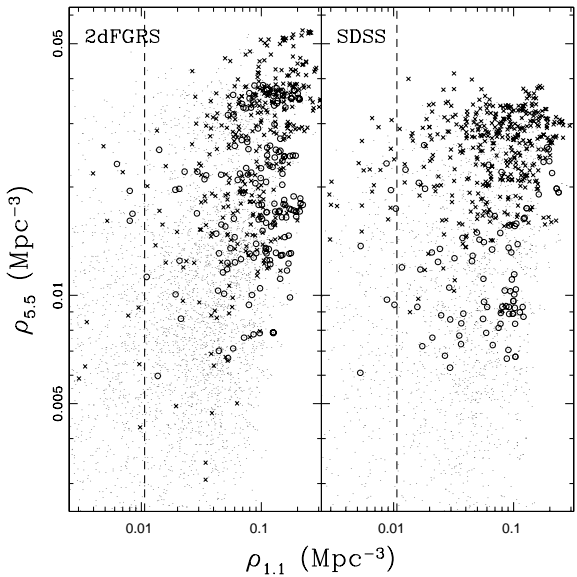}
\caption{Galaxies in the 2dFGRS  {\it (left panel)} and SDSS {\it (right
    panel)} are shown as a function of density computed on 1.1 Mpc and
  5.5 Mpc scales, using a Gaussian filtering kernel ({\it small dots}).  Galaxies in groups with
  $\sigma=$200--400$\kms$ are shown as {\it open circles}, while those
  with $\sigma>600\kms$ are represented as {\it crosses}.  Only galaxies
  that are at least 11 Mpc from a survey boundary are considered.  The
  {\it dashed line} shows the minimum reliable density; below this
  there are fewer than one galaxy within the filter size $\theta$
  (the corresponding limit on $\rho_{5.5}$ is off the scale of the plot).
\label{fig-g11g55}}
\end{figure}
We have shown that the fraction of star-forming galaxies depends
strongly on the local density, measured on scales $\lesssim$ 1 Mpc, and
that this correlation is approximately independent of the velocity
dispersion of the embedding cluster or group.  However, velocity
dispersion may not be the best measurement of large-scale structure; in
this subsection, we will 
compare densities measured on $1.1$ Mpc and $5.5$ Mpc scales,  to determine whether
\ewha\ shows any independent sensitivity to structure on large-scales.

In Fig.~\ref{fig-g11g55} we compare
the Gaussian-filtered densities $\rho_{1.1}$ and $\rho_{5.5}$ for
galaxies in the SDSS and 2dFGRS  samples.  Galaxies are only considered
if they are at least 11 Mpc from a survey boundary.  In
Appendix~\ref{sec-kde} we show that these density estimators are good
tracers of intuitively dense and low-density regions, despite
the complications of peculiar velocities in clusters
(Fig.~\ref{fig-posden}).  Note that the 2dFGRS data extend to larger
values of $\rho_{5.5}$ than the SDSS; these data arise from a large
supercluster region at $z\sim 0.0839$ and $\alpha\sim195.5$, $\delta\sim-2.9$ (J2000
degrees).  Fig.~\ref{fig-g11g55} shows a clear separation
between galaxies in groups ($200\kms<\sigma<400\kms$) and clusters
($\sigma>600\kms$).  
Galaxies in both environments span a similar range in local
environment, characterised by $\rho_{1.1}$.  However, galaxies in the high-velocity
dispersion clusters lie at higher densities on 5.5 Mpc
scales, as expected since they are physically larger systems.  This is particularly true for the SDSS sample,
where the groups are selected to have approximately Gaussian
distributed velocities, so that there is likely to be less scatter in the
relation between $\sigma$ and virial mass.  Note that the higher velocity dispersions of clusters means
that this pseudo-three-dimensional measurement will {\it underestimate}
the density, more so than in groups.  Thus the real difference between
the clusters and groups is even larger than shown here.
\begin{figure*}
\leavevmode \epsfysize=8cm \epsfxsize=8cm \epsfbox{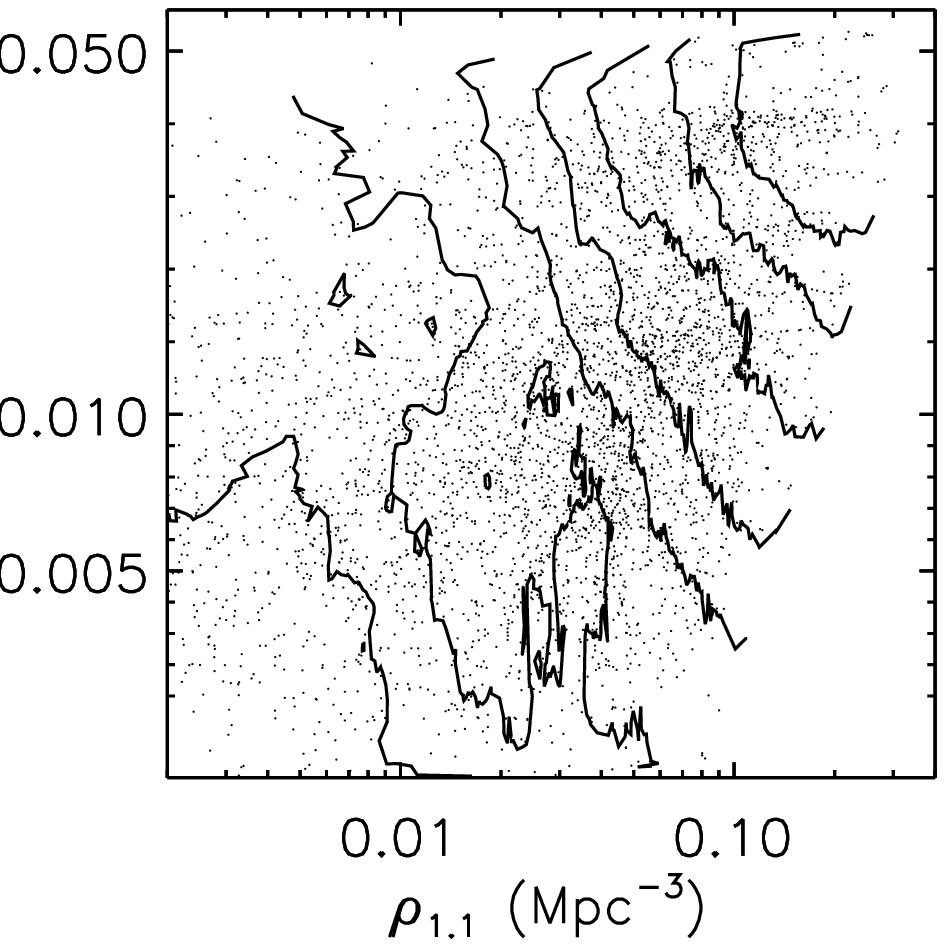}\hskip 1cm
\epsfysize=8cm \epsfxsize=8cm \epsfbox{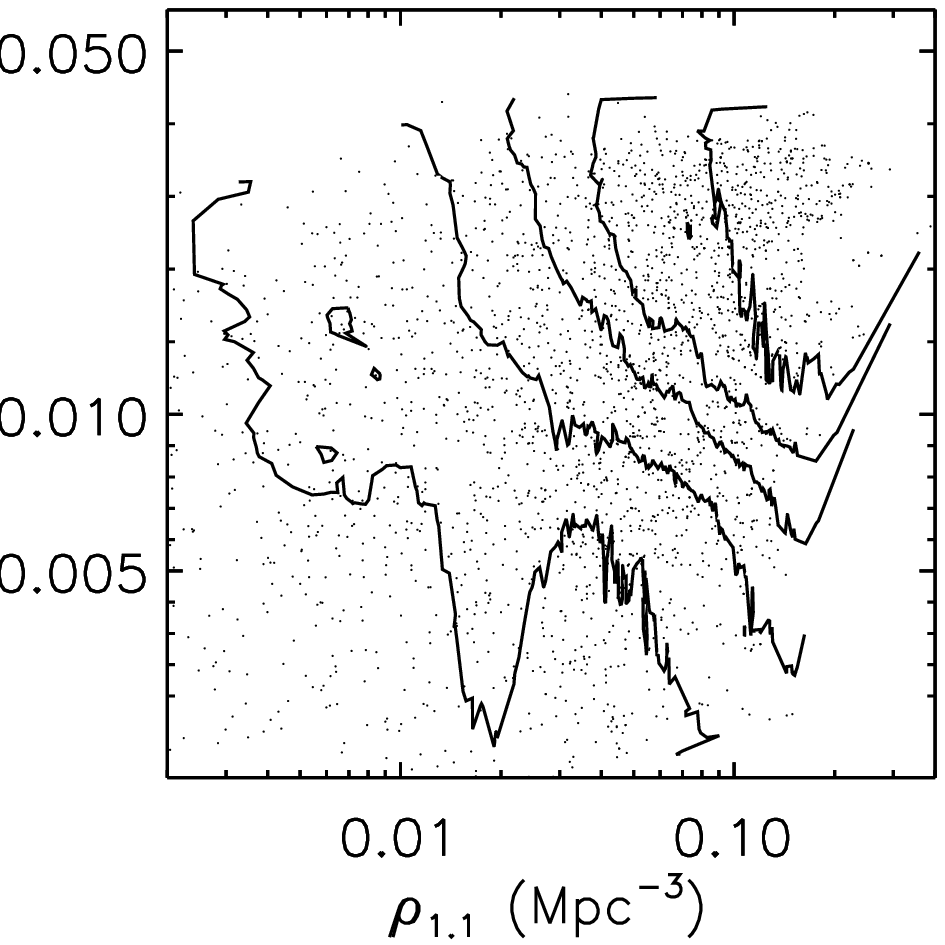}
\caption{Galaxies in the 2dFGRS  {\it (left panel)} and SDSS {\it (right
    panel)} are shown as a function of density computed on 1.1 Mpc and
  5.5 Mpc scales, using a Gaussian filtering kernel.  Only galaxies
  that are at least 11 Mpc from a survey boundary are considered.  The
  contours trace the fraction of galaxies with \ewha$>4$\AA, computed for the nearest 500
  galaxies at each point in this plane.  The contours are spaced in
  steps of 0.05, and increase toward lower densities.   For
the SDSS sample the contours span fractions 0.25--0.45, while for the 2dFGRS they
span 0.35--0.65. 
\label{fig-hacont}}
\end{figure*}

In Fig.~\ref{fig-hacont}, we illustrate how the fraction of
emission-line galaxies
(measured from the nearest 500 galaxies in this plane)
depends on these two very different density scales.  Interestingly, the
contours are not parallel with either axis, which indicates that the
population composition shows a
dependance on large ($>5$ Mpc) scales, in addition to
the more local density measured at 1.1 Mpc.  That is, the fraction of
emission line galaxies is lower in regions that are overdense on 5.5
Mpc scales, even when the local ($1.1$ Mpc) overdensity is the same.
We have checked that this is not an artifact 
of the smoothing or the correlation between $\rho_{1.1}$ and
$\rho_{5.5}$, in the following way.  For each galaxy we assign a value
of $\rho_{5.5}$, chosen at random from among galaxies that have similar
(within 10\%) values of $\rho_{1.1}$.
This preserves the correlations between $\rho_{1.1}$ and $\rho_{5.5}$,
as well as between $\rho_{1.1}$ and \ewha, but removes any residual
correlation between $\rho_{5.5}$ and \ewha.  In this case, the contours
lie nearly parallel to the $\rho_{5.5}$ axis, confirming that the
correlation we see in 
Fig.~\ref{fig-hacont} is real.  There is some evidence that, at high
densities, the fraction of H$\alpha$--emitting galaxies is mostly
dependent on $\rho_{1.1}$, and the dependence on larger scales becomes
more important at lower densities.
The dependence on large-scale densities is evidently stronger than the
dependence on velocity dispersion (Fig.~\ref{fig-haden}), despite the
correlation between them (Fig.~\ref{fig-g11g55}).  

A cautionary note needs to be added, however.  Since $\rho_{1.1}$ and
$\rho_{5.5}$ are intrinsically correlated, errors on these measurements
can give rise to the trends shown in Fig.~\ref{fig-hacont}, even if the
galaxy population only depends on one parameter.  However, a comparison
with mock catalogues strongly suggests that  the observed dependence on
both density scales is real (Balogh et al., in prep.). 

\section{Discussion}\label{sec-discuss}
\subsection{Overview}
Thanks to the unique size and homogeneity of the 2dFGRS and SDSS
datasets, we have been able to trace the H$\alpha$ distribution of
galaxies over the full range of environments at the present day.  This
analysis has revealed three new, important clues about the process of
galaxy formation:
\begin{itemize}
\item[1.] The change in \ewha\ distribution as a function of
environment is predominantly due to a change in the relative number of
star--forming and non-star-forming galaxies; the \ewha\ distribution of actively star-forming
galaxies themselves do not appear to depend strongly on environment.
This is a surprising result which challenges theories in which the
environment at the present day induces a transformation in galaxy
properties, as we discuss in detail in Section~\ref{sec-colours}.

\item[2.] The fraction of star--forming
galaxies increases continuously with decreasing density.   However, even
at the lowest densities there is a substantial fraction of galaxies
with negligible H$\alpha$ emission; we will discuss the consequences of
this in Section~\ref{sec-isol}.  There is 
evidence that the correlation with environment becomes steeper at
$\Sigma_5\gtrsim 1$ Mpc$^{-2}$, an effect that is also seen in the
morphology--density relation \citep{PG84,Treu03}.  This may represent
an interesting scale at which different physics 
becomes important; for example, as noted by \citet{PG84}, at this
density the
typical dynamical time of the group is longer than a Hubble time, so galaxy-galaxy
interactions may start to play a role \citep[see also][]{ZM98,ZM00}.  

\item[3.] The fraction of galaxies with H$\alpha$ emission may depend not only
  on the local density, but also on the density on much larger scales,
  $\sim 5.5$ Mpc.  If this trend is real, it means the galaxy population
  must be only {\it indirectly} related to its present-day
  environment.  We discuss the implications of this in
  Section~\ref{sec-lss}.
\end{itemize}

These three clues allow us to make advances in our understanding of the
 evolution of star formation in
the Universe (Section~\ref{sec-madau}), and of any 
additional physics that may take place in dense environments
(Section~\ref{sec-physics}).

\subsection{Star-forming galaxies in dense environments}\label{sec-colours}
If galaxies at the present day are evolving as they move from
low-density regions to high-density regions, we should see a signature
of this transformation which depends on the relevant timescale.  In
particular, if
all galaxies experience a gradual ($\gtrsim 2 $ Gyr) decline in star
formation when they become bound to groups or clusters \citep{infall,Bekki02},
we expect star-forming galaxies in dense regions to show
  systematically lower  \ewha.  
Fig.~\ref{fig-half} shows that this is not the case,
implying that such
a gradual decline is not a common phenomenon at the present time.

On the other hand, if the SFR declines rapidly to zero, this might not be
reflected in Fig.~\ref{fig-half}, since galaxies will quickly move from
the star-forming distribution to the quiescent one with \ewha$<4$\AA.  
Instead, we need to look at a
longer-lived tracer of star formation (i.e. one which changes slowly
after star formation ceases) to observe this effect.  For example, the
H$\delta$ absorption line, which remains strong for $\sim 1$ Gyr after
star formation ceases, is a common diagnostic of recent activity.   The
extreme  rarity ($<0.1$ per cent) of bright galaxies with strong H$\delta$
but negligible star formation at low redshift provides  evidence against the
possibility that SFR has recently declined on a short timescale for a
significant number of galaxies \citep{Z+96,tomo-EA1}; however, this
depends on an accurate quantification of the duration of the enhanced
H$\delta$ phase, which is systematically uncertain to within at least a factor
of two.  In particular, if star formation is truncated in normal
spirals, without a preceding burst, the H$\delta$ line never gets very
strong, and is only enhanced for a short time \citep[e.g.][]{PSG}.

\begin{figure}
\leavevmode \epsfysize=8cm \epsfbox{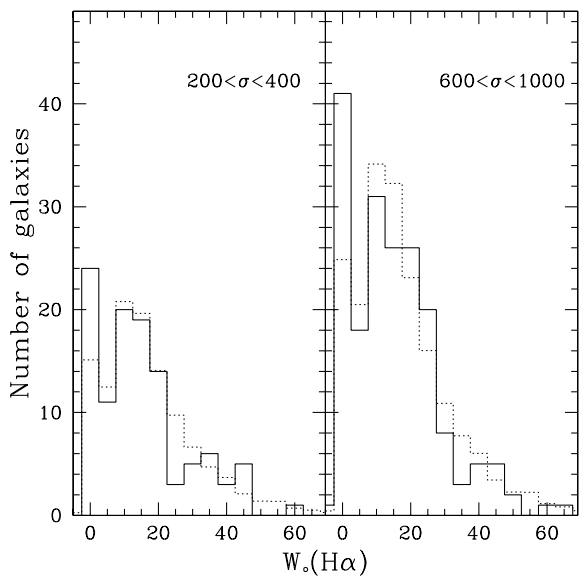}
\caption{The \ewha\ distribution for blue galaxies ($(g-r)_0<0.7$) in the SDSS
  volume-limited sample.  The {\it solid line} shows the distribution
  for galaxies in groups {\it(left panel)} and clusters {\it (right
    panel)}, while the {\it dotted histogram} shows the distribution
  for all blue galaxies in the SDSS sample, renormalised to include the
  same  area as the solid lines.  
\label{fig-colSFR}}
\end{figure}

In a similar way, we can use the accurate colours of the SDSS, which do
not suffer from aperture bias, to trace the longer lived stellar
population. In particular, we will consider 
the rest frame Petrosian $(g-r)$ colour, denoted $(g-r)_0$,  and use the
k-corrections of \citet{Blanton03}. 
To isolate the population of galaxies with recent star formation,
we select blue galaxies as those with $(g-r)_0<0.7$, which will include
any galaxy that has formed stars within the last $\sim 300$ Myr, using
the latest \citet{BC03} stellar population models\footnote{Based on
  a model with a Salpeter initial mass function,
and a SFR which declines exponentially ($\tau=4$ Gyr) for 10 Gyr.}.
Fig.~\ref{fig-colSFR} shows the \ewha\ distribution
for all such blue galaxies, compared with that of blue galaxies in groups ($\sigma=200$--$400\kms$) and clusters
($\sigma>600\kms$).  All three distributions are consistent with being
drawn from the same parent population.  If cluster galaxies had
slowly declining SFRs they would remain blue, but with reduced \ewha\
\citep[e.g][]{S+02}, contrary to our observations.
This reinforces our conclusions
drawn from Fig.~\ref{fig-half}, that star-forming galaxies
in dense environments have normal SFRs, and are not being
inhibited by their environment at the present day.  

However, we are still not able to rule out the possibility that the
environment has caused the SFR to
decline {\it rapidly} in a 
substantial number of cluster galaxies, as we now demonstrate.  There
are  617 galaxies with \ewha$<4$\AA\ in the  $\sigma>600\kms$ 
clusters.
If the clusters have been constantly, and
uniformly, truncating star formation in active galaxies over the last
$\sim 10$ Gyr, we would only need to
find $\sim 20$ blue galaxies with \ewha$<4$\AA\ in clusters with
$\sigma>600\kms$ at the present day
to account for these currently inactive galaxies.  
This number is not inconsistent with Fig.~\ref{fig-colSFR}; in fact there is
perhaps evidence for such a population, though this may equally well be
contamination from the tail of the distinct, red galaxy population at
$(g-r)_0>0.7$.  By the same argument, our results are consistent with
a model in which 20 per cent of the cluster members not presently forming
stars had their star formation truncated sometime within the last 2 Gyr.

Thus, the $(g-r)_0$ colour is too sensitive to recent star
formation for us to put strong constraints on the number of cluster
galaxies which have had their star formation truncated in the last few
Gigayears.  Galaxy morphology may provide a better indicator, since
an observable disk structure should persist for $\gtrsim 1$ Gyr after star
formation ceases \citep[e.g.][]{Bekki02}.  Past 
analysis (Balogh et al. 1998\nocite{B+98}; Hashimoto et
al. 1998\nocite{H+98}; Paper~II; Paper~I;
Girardi \etal\ 2003\nocite{Girardi03}) has indeed suggested that
morphology and SFR are partly independent.  Furthermore, the existence
in clusters of HI-deficient spirals \citep[e.g][]{Solanes01}, red, spiral galaxies
with little star-formation \citep[e.g.][]{P+99,lowlx-spectra_short} and spiral
galaxies with unusually smooth structure \citep{McIntosh03} all argue
for some sort of transformation of spiral galaxies to be taking place
in clusters.  However, many of these results are
based on a fairly coarse morphological binning; therefore, we reserve drawing
firm conclusions until reliable, automated morphological measurements
are available for the present samples \citep[e.g.][]{Liske,KM03,Blanton03BB}.

\subsection{Isolated galaxies}\label{sec-isol}
Although the fraction of emission-line galaxies continually increases
with decreasing density, it never gets much larger than $\sim 70$ per cent, even in the lowest density
environments studied here.  In these empty regions of
the Universe, environment is not likely to have played a large role in
galaxy evolution; therefore many galaxies must have ceased their star-formation
activity for reasons independent of their surroundings (but see below).  In
Fig.~\ref{fig-isol} we show the \ewha\ distribution for 
galaxies in the lowest-density environments, with $\rho_{1.1}<0.01$
Mpc$^{-3}$ and $\rho_{5.5}<0.005$~Mpc$^{-3}$, and unassociated with any
catalogued group or cluster.  We combine all galaxies
from both the 2dFGRS and SDSS samples, and consider only those
galaxies which lie at least 11 Mpc from a survey boundary.  Even in these extremely
sparse regions of space, only $\sim 70$ per cent of galaxies are
forming a significant number of stars, with \ewha$>4$\AA, and this fraction is similar
for both faint and bright galaxies\footnote {Note that only
the brightest isolated galaxies  are sure to have no companions of
comparable brightness, since fainter galaxies could have
neighbours that are just below our luminosity limit.} (shown separately in Fig.~\ref{fig-isol}).
\begin{figure}
\leavevmode \epsfysize=8cm \epsfbox{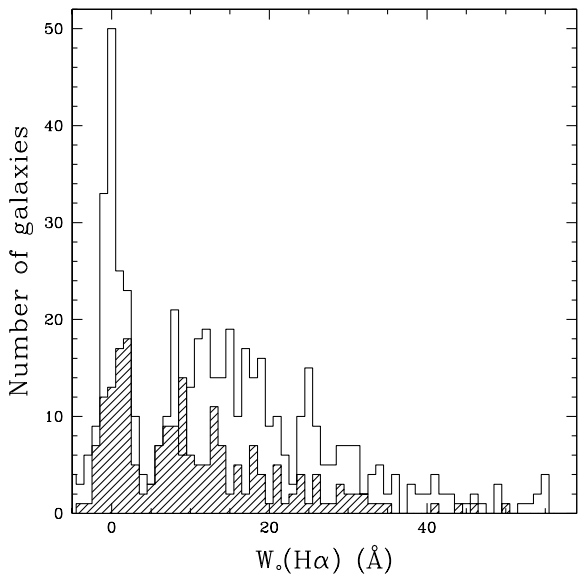}
\caption{The \ewha\ distribution for all SDSS and 2dFGRS galaxies in low density
  environments ($\rho_{1.1}<0.01$~Mpc$^{-3}$ and
  $\rho_{5.5}<0.005$~Mpc$^{-3}$, unassociated with a group or cluster).  
The {\it shaded histogram} shows
  the distribution for bright ($>L^\ast$) galaxies, and the {\it solid
    line} represents the fainter population. 
\label{fig-isol}}
\end{figure}
A possible interpretation is that the $\sim 30$ per cent of bright, isolated
galaxies with no sign of star formation are fossil groups
\citep{PBnat,MZ99,SNDS,Jones03}.  Although we await morphological
confirmation, we note that these galaxies do have the colours typical
of elliptical galaxies, $(g-r)_0\sim0.75\pm0.05$.  In this case, their
isolation may be a misleading representation of environment; in fact,
they may represent the most dense environments, where all bright,
surrounding galaxies have merged into one.  

\subsection{Dependence on large-scale structure}\label{sec-lss}
We showed in \S~\ref{sec-lgd} that, although the galaxy population depends strongly on its local
environment,  it may also be sensitive to the density on
$\gtrsim 5$ Mpc scales, especially at low densities.  That is, fewer galaxies in supercluster-like environments have
significant H$\alpha$ emission, relative to galaxies in environments
with similar local densities.  We currently treat this conclusion as
tentative, because a similar trend could be caused by
uncertainties in the correlated densities $\rho_{1.1}$ and
$\rho_{5.5}$; however, comparison with mock catalogues strongly suggest
this is not the case (Balogh et al., in prep).  A 
similar dependence on large-scale structure has previously been
observed in the morphologies of galaxies in clusters at $z\sim
0.2$ \citep{lowlx-morph_short}.  Using high resolution {\it Hubble
  Space Telescope} imaging to decompose the disk and bulge components of
cluster members, \citet{lowlx-morph_short} found evidence that the
bulges of galaxies at a fixed local density are systematically brighter
in massive clusters (as traced by their X-ray 
emission) than in low mass clusters.  On the other hand, the disk luminosity function does
not show this dependence.  It was suggested that bulge components grow
preferentially within large-scale overdensities, perhaps due to richer
merger histories, but that the disk properties are not sensitive to
this structure.

However, our results show that the galaxy population depends more
strongly on the large-scale density than on the mass of the embedding halo,
as evidenced by the lack of correlation with cluster velocity
dispersion, at a fixed $\Sigma_5$ (Fig.~\ref{fig-haden}).  
This, together with the results of Section~\ref{sec-colours}, suggests that galaxy
properties are only indirectly related to their environment at the present
day.  For example, the early-type population that dominates clusters
today likely arises from the fact that galaxies in dense regions form
earlier, and experience more galaxy-galaxy interactions throughout
their longer lifetime, than galaxies in underdense regions
\citep{ZM98,BCOS,BDS}.  Another possibility is that galaxy evolution is
strongly affected by tidal forces, 
which arise from structure on all scales \citep[e.g.][]{Gnedin03a,MW00}.

\subsection{Implications for the global SFR evolution}\label{sec-madau}
It has been proposed that changes in the typical environment, due to
the hierarchical growth of structure, drive the evolution in global SFR
\citep{BB02,BalBow03}.  This hypothesis maintains that the correlation between SFR and
local density remains unchanged, but galaxies at higher redshift
typically lie in lower density environments.  However, this interpretation is inconsistent with the present data, as
we demonstrate here.
The fraction of star--forming galaxies (\ewha$>4$\AA) is $\sim0.4$ in the SDSS and $\sim0.57$
in the 2dFGRS.  This is close to the maximum fraction achieved in any
environment at $z=0$ (e.g. Figs.~\ref{fig-haden} and \ref{fig-kde};
also Section~\ref{sec-isol}).
On the other hand, the average SFR at $z\sim 0.4$ is expected to be about
75\% higher than at the present day.  This can be inferred from
the increase in the ultraviolet luminosity density of the Universe \citep[e.g.,][]{Wilson+02};
a similar result is obtained by comparing the average [O{\sc ii}] equivalent width at
$z\sim 0.3$ from the CNOC surveys with that at $z=0$ (Paper~II).  To
achieve such a dramatic increase in the mean SFR, either the fraction of star-forming
galaxies at $\sim 0.4$ is much higher than anywhere at the present day, or the
typical star-forming galaxy has a higher SFR.  It is interesting that \citet{Treu03} find the fraction of
morphologically--classified early-type galaxies in low-density regions
has not evolved substantially between $z=0.4$ and $z=0$, suggesting
that evolution is of the latter type.  Either way, the galaxy
population at $z=0.4$ must be different from that in any environment at
$z=0$.  

A direct test of evolution in the SFR--density correlation is
not yet possible, as intermediate- and high-redshift cluster data do not generally
extend far enough from the cluster
centre \citep{B+97,C+01,lowlx-spectra_short,A1689}.  
\citet{Kodama_cl0939}, however, have analysed
multicolour Subaru data which
shows that the red sequence in the $z\sim 0.4$ cluster Cl0939 first becomes apparent in regions where
the local projected density exceeds 30 Mpc$^{-2}$ \citep[see the
Erratum of][]{Kodama_err}.  
Their photometric observations
are much deeper than ours, and are complete to $\sim 0.02L_V^\ast$.  Correcting for this
difference, their measurement of the critical density corresponds to
$\Sigma_5\sim4\pm2$ Mpc$^{-2}$ in our units.  This is similar to the
result found in Paper~I and Paper~II, and seen in Fig.~\ref{fig-sfrden1}, which
suggests that any evolution in the SFR--density relation has been
small.  However, any serious interpretation of this comparison should
wait until comparable, spectroscopic measurements can be made of the
galaxy population composition as a continuous function of density.

\subsection{The physics of galaxy evolution}\label{sec-physics}
Based on the results presented here, it seems unlikely that a substantial fraction of the
star-forming galaxy population is today undergoing a physical
transformation induced by its environment.  Instead, the observed trends with density at $z\sim
0$ are probably only related indirectly to their environment, and
the physics which
determines the final composition of galaxy groups and clusters 
probably took place at a much earlier time.   Similar conclusions were
reached by \citet{ZM98}, based on the high fractions of early-type
galaxies in nearby, X-ray groups.  Observations of the local
Universe alone can only put weak constraints on what the relevant physics might
be, except to say that it is unlikely to be dominated by 
ram-pressure stripping \citep[e.g.][]{GG,QMB}.  This mechanism has the distinct
advantage that there is good observational evidence for such activity
\citep[e.g.][]{Veilleux,Vollmer00,Gavazzi02,vanGorkom03}; however, it seems
unlikely that it could play any role in the lowest-density environments.

Similarly, it may be difficult to interpret our results in the context
of models in which SFR declines slowly in galaxies that are accreted
into groups and clusters \citep{infall,Diaferio,Bekki02}, for reasons discussed
in Section~\ref{sec-colours}.  In particular, galaxy groups are
  expected to represent the first level of the hierarchy in which
  heating and stripping of the hot halo gas is likely to influence
  star formation \citep[e.g.][]{Diaferio,Cole2000,BCOS,HS03}, yet blue
  galaxies within such groups appear to have a normal \ewha\
  distribution, suggesting normal, recent star formation histories.
However, the predictions of hierarchical models are not straightforward, because
local density, halo mass, and formation time are all correlated in a
non-trivial way.  It is possible that including the effects of infall
\citep{Erica} and projection effects \citep{Diaferio}, in addition to a
slow-decay model \citep{infall} conspire to keep the shape of the
\ewha\ distribution constant.  We will therefore leave a detailed comparison with
theory for a subsequent paper (Balogh et al., in prep).

A more viable explanation is perhaps suggested by observations of close galaxy pairs, which
are the only environmentally--selected
population to show {\it enhancements} of star formation
\citep{Barton,Lambas}.  These bursts likely lead to the rapid
consumption of cold gas, and the eventual formation of gas-poor
elliptical galaxies with little star formation \citep[e.g.][]{PBnat,MZ99,SNDS}.
Such galaxy-galaxy interactions must typically occur long before a galaxy
is bound to a virialised group or cluster, since the SFRs in those
environments are so low \citep[e.g.][]{ZM98}.  In fact, there is
evidence for interaction--induced star formation in the unvirialised
regions of clusters \citep[e.g.][]{CR97,MW00}, and for a correlation
between morphological distortions and environment \citep{HO00}.
The observed correlation with local density then arises
because galaxies in dense regions have typically had more
interactions, over a longer period of time, than those in low-density
regions \citep[e.g.][]{BDS,Gnedin03a,Gnedin03b}.  Since the interaction rate may increase substantially with
redshift \citep{Patton,C+03},  we would expect most of
the environmentally-induced galaxy transformation to have taken place at higher
redshift; this would be consistent with the strong evolution
observed in the fraction of
post-starburst galaxies \citep{PoggEA03}.

We might therefore be
surprised that the correlation between emission--line fraction and
local density is so similar in the groups in the present study,
relative to other environments,  since the low velocity
dispersions and high densities in groups should encourage
interactions \citep[e.g.][]{AF80,Barnes85,Merritt85}.  Our groups are very heterogeneous, and not all
are necessarily bound systems.  It is possible that interactions are
important in specific types of group, such as compact groups.  We will
leave such investigations for future work.

Finally, we note that there is no observed correlation between the
fraction of AGN and environment at the present day \citep{Miller_agn}.
This is puzzling, especially in the context of models where star
formation and AGN activity are both linked to the availability of cold
gas \citep{KH,Granato03,DiMatteo03}.  It appears that star formation activity must
stop (or become undetectable) when there is still some cold gas left.
One possibility is that the gas density drops below
a certain threshold necessary for star formation \citep[e.g.][]{K89,MK01,WB02,BPBG,Gnedin03a}, but
can still flow to the centre and fuel an AGN.

\section{Conclusions}\label{sec-conc}
We have made a joint analysis of 24968 galaxies selected from the SDSS
and 2dFGRS.  The distribution of \ewha\ among these galaxies is
bimodal, consisting of an active population of galaxies with
\ewha$>4$ \AA, and a quiescent population with no significant star
formation at the present day.  We have investigated how this
distribution depends on environment, as characterised by:
\begin{itemize}
\item The projected surface number density of galaxies $\Sigma_5$, determined
  from the distance to the fifth-nearest neighbour.
\item The three-dimensional number density of galaxies measured on
  1.1 Mpc and 5.5 Mpc scales, $\rho_{1.1}$ and $\rho_{5.5}$.
\item The velocity dispersion of the embedding structure, as
  determined from the group catalogues of \citet{Eke-groups} and
  Nichol, Miller et al. (in preparation).
\end{itemize}
We use these different measures of environment to establish the scales
and structures on which the present-day galaxy population depends.  Our
findings are summarised as follows:

\begin{itemize}
\item[1.] The distribution of H$\alpha$ line strength for the star-forming
  population, selected on \ewha\ or $(g-r)$ colour, does not itself
  depend strongly on environment.  Thus, it is unlikely that SFRs are gradually
  decreasing in a substantial number of star-forming galaxies in or
  near dense environments today. 
\item[2.] The fraction of galaxies with \ewha$>4$\AA\ decreases steadily
  with increasing local density.  There is evidence that it decreases
  more strongly at densities exceeding $\Sigma_5\sim1$ Mpc$^{-2}$, or
  $\rho_{1.1}\gtrsim 0.05$ Mpc$^{-3}$.  The persistence of this
  correlation at low densities means that ram--pressure stripping, at
  any redshift, cannot be the only  physical  mechanism at work.
\item[3.] The fraction of galaxies brighter than $M^\ast+1$ with \ewha$>4$\AA\ is never more than
  $\sim 70$ per cent, even in the least dense environments explored
  here.  We have shown that this means the recent decline in
  globally-averaged star formation rate cannot be 
wholly due to the growth of large scale structure.
\item[4.] The emission-line
  fraction of a galaxy population appears to depend both on the local environment
  (on $\sim 1$ Mpc scales) {\it and} on the large-scale structure as
  parameterised by the density
  $\sim 5.5$ Mpc scales.  
  There is little further dependence
   on the velocity dispersion of the group or cluster in
  which the galaxy is embedded.  This result suggests that the composition of the galaxy population today is likely
  related {\it indirectly} to its present environment \citep[see also][]{ZM98,ZM00}.
\end{itemize}

The most likely physical explanations for the correlation between
\ewha\ distribution and environment at $z=0$ are those which are
effective over a large range of environment, affect the SFR on short
($<1$ Gyr) timescales, and were much more
effective in the past.  As suggested by previous authors
\citep[e.g.][]{Z+96,MZ98,H+98,HO00}, one good candidate is starbursts induced by
galaxy interactions, since close pairs of galaxies are the only
environment known to directly provoke a physical transformation.  Such
interactions will likely be more common at high redshift, and will have
had more time to influence galaxies that end up in high density environments.

\section*{Acknowledgements}
We gratefully acknowledge the efforts of all persons whose
contributions led to
the success of the 2dFGRS and SDSS, which in turn have made this work possible.
In addition, MLB acknowledges helpful discussions with Ann Zabludoff, Tadayuki Kodama,
Masayuki Tanaka and Diego Lambas; also, he thanks Carolyn M$^{\rm
  c}$Coey for a careful reading of the manuscript which substantially
improved its clarity.
MLB and RGB acknowledge financial support from PPARC fellowships,
numbered PPA/P/S/2001/00298 and PPA/Y/S/2001/00407, respectively.
WJC acknowledges the financial support of the Australian
Research Council throughout the course of this work.

\setlength{\bibhang}{2.0em}

\bibliography{ms}

\begin{thebibliography}{127}
\expandafter\ifx\csname natexlab\endcsname\relax\def\natexlab#1{#1}\fi

\bibitem[{{Aarseth} \& {Fall}(1980)}]{AF80}
{Aarseth}, S.~J. \& {Fall}, S.~M. 1980, ApJ, 236, 43

\bibitem[{{Abazajian} {et~al.}(2003)}]{DR1}
{Abazajian}, K. {et~al.} 2003, AJ, 126, 2081

\bibitem[{{Afonso} {et~al.}(2003){Afonso}, {Hopkins}, {Mobasher}, \&
  {Almeida}}]{Afonso}
{Afonso}, J., {Hopkins}, A., {Mobasher}, B., \& {Almeida}, C. 2003, ApJ, 597,
  269

\bibitem[{{Allington-Smith} {et~al.}(1993){Allington-Smith}, {Ellis}, {Zirbel},
  \& {Oemler}}]{A+93}
{Allington-Smith}, J.~R., {Ellis}, R., {Zirbel}, E.~L., \& {Oemler}, A.~J.
  1993, ApJ, 404, 521

\bibitem[{{Baldry} {et~al.}(2003){Baldry}, {Glazebrook}, {Brinkmann},
  {Ivezi\'{c}}, {Lupton}, {Nichol}, \& {Szalay}}]{Baldry03}
{Baldry}, I.~K., {Glazebrook}, K., {Brinkmann}, J., {Ivezi\'{c}}, Z., {Lupton},
  R.~H., {Nichol}, R.~C., \& {Szalay}, A.~S. 2003, ApJ, 600, in
  press,astroph/0309710

\bibitem[{{Balland} {et~al.}(2003){Balland}, {Devriendt}, \& {Silk}}]{BDS}
{Balland}, C., {Devriendt}, J.~E.~G., \& {Silk}, J. 2003, MNRAS, 343, 107

\bibitem[{{Balogh} {et~al.}(2002{\natexlab{a}}){Balogh}, {Bower},
  {et~al.}}]{lowlx-spectra_short}
{Balogh}, M., {Bower}, R.~G., {et~al.} 2002{\natexlab{a}}, MNRAS, 337, 256

\bibitem[{{Balogh} \& {Bower}(2003)}]{BB02}
{Balogh}, M.~L. \& {Bower}, R.~G. 2003, in Revista Mexicana de Astronomia y
  Astrofisica Conference Series, 220--221

\bibitem[{{Balogh} {et~al.}(2002{\natexlab{b}}){Balogh}, {Couch}, {Smail},
  {Bower}, \& {Glazebrook}}]{A1689}
{Balogh}, M.~L., {Couch}, W.~J., {Smail}, I., {Bower}, R.~G., \& {Glazebrook},
  K. 2002{\natexlab{b}}, MNRAS, 335, 10

\bibitem[{Balogh {et~al.}(1997)Balogh, Morris, Yee, Carlberg, \&
  Ellingson}]{B+97}
Balogh, M.~L., Morris, S.~L., Yee, H. K.~C., Carlberg, R.~G., \& Ellingson, E.
  1997, ApJL, 488, 75

\bibitem[{{Balogh} {et~al.}(1999){Balogh}, {Morris}, {Yee}, {Carlberg}, \&
  {Ellingson}}]{PSG}
{Balogh}, M.~L., {Morris}, S.~L., {Yee}, H.~K.~C., {Carlberg}, R.~G., \&
  {Ellingson}, E. 1999, ApJ, 527, 54

\bibitem[{{Balogh} {et~al.}(2000){Balogh}, {Navarro}, \& {Morris}}]{infall}
{Balogh}, M.~L., {Navarro}, J.~F., \& {Morris}, S.~L. 2000, ApJ, 540, 113

\bibitem[{{Balogh} {et~al.}(1998){Balogh}, {Schade}, {Morris}, {Yee},
  {Carlberg}, \& {Ellingson}}]{B+98}
{Balogh}, M.~L., {Schade}, D., {Morris}, S.~L., {Yee}, H.~K.~C., {Carlberg},
  R.~G., \& {Ellingson}, E. 1998, ApJL, 504, L75

\bibitem[{{Balogh} {et~al.}(2002{\natexlab{c}}){Balogh}, {Smail}, {Bower},
  {et~al.}}]{lowlx-morph_short}
{Balogh}, M.~L., {Smail}, I., {Bower}, R.~G., {et~al.} 2002{\natexlab{c}}, ApJ,
  566, 123

\bibitem[{{Barnes}(1985)}]{Barnes85}
{Barnes}, J. 1985, MNRAS, 215, 517

\bibitem[{{Barton} {et~al.}(2000){Barton}, {Geller}, \& {Kenyon}}]{Barton}
{Barton}, E.~J., {Geller}, M.~J., \& {Kenyon}, S.~J. 2000, ApJ, 530, 660

\bibitem[{{Baugh} {et~al.}(1999){Baugh}, {Benson}, {Cole}, {Frenk}, \&
  {Lacey}}]{Baugh99}
{Baugh}, C.~M., {Benson}, A.~J., {Cole}, S., {Frenk}, C.~S., \& {Lacey}, C.~G.
  1999, MNRAS, 305, L21

\bibitem[{{Beers} {et~al.}(1990){Beers}, {Flynn}, \& {Gebhardt}}]{Beers}
{Beers}, T.~C., {Flynn}, K., \& {Gebhardt}, K. 1990, AJ, 100, 32

\bibitem[{{Bekki} {et~al.}(2002){Bekki}, {Couch}, \& {Shioya}}]{Bekki02}
{Bekki}, K., {Couch}, W.~J., \& {Shioya}, Y. 2002, ApJ, 577, 651

\bibitem[{{Benson} {et~al.}(2003){Benson}, {Hoyle}, {Torres}, \&
  {Vogeley}}]{Benson_voids}
{Benson}, A.~J., {Hoyle}, F., {Torres}, F., \& {Vogeley}, M.~S. 2003, MNRAS,
  340, 160

\bibitem[{{Blanton} {et~al.}(1999){Blanton}, {Cen}, {Ostriker}, \&
  {Strauss}}]{BCOS}
{Blanton}, M., {Cen}, R., {Ostriker}, J.~P., \& {Strauss}, M.~A. 1999, ApJ,
  522, 590

\bibitem[{{Blanton} {et~al.}(2003{\natexlab{a}}){Blanton}, {Hogg},
  {et~al.}}]{Blanton03}
{Blanton}, M.~R., {Hogg}, D.~W., {et~al.} 2003{\natexlab{a}}, ApJ, 592, 819

\bibitem[{{Blanton} {et~al.}(2003{\natexlab{b}})}]{Blanton03BB}
{Blanton}, M.~R. {et~al.} 2003{\natexlab{b}}, ApJ, 594, 186

\bibitem[{{Boissier} {et~al.}(2003){Boissier}, {Prantzos}, {Boselli}, \&
  {Gavazzi}}]{BPBG}
{Boissier}, S., {Prantzos}, N., {Boselli}, A., \& {Gavazzi}, G. 2003, MNRAS,
  346, 1215

\bibitem[{{Bower} \& {Balogh}(2003)}]{BalBow03}
{Bower}, R.~G. \& {Balogh}, M.~L. 2003, in Carnegie Observatories Astrophysics
  Series, Vol. 3, Clusters of Galaxies: Probes of Cosmological Structure and
  Galaxy Evolution, ed. A.~D. J.S.~Mulchaey \& A.~Omeler, astro--ph/0306342

\bibitem[{{Bruzual} \& {Charlot}(2003)}]{BC03}
{Bruzual}, G. \& {Charlot}, S. 2003, MNRAS, 344, 1000

\bibitem[{{Caldwell} \& {Rose}(1997)}]{CR97}
{Caldwell}, N. \& {Rose}, J.~A. 1997, AJ, 113, 492

\bibitem[{{Carlberg} {et~al.}(2001){Carlberg}, {Yee}, {Morris},
  {et~al.}}]{CNOC_groups2_short}
{Carlberg}, R.~G., {Yee}, H.~K.~C., {Morris}, S.~L., {et~al.} 2001, ApJ, 563,
  736

\bibitem[{{Charlot} \& {Longhetti}(2001)}]{CL}
{Charlot}, S.~. \& {Longhetti}, M. 2001, MNRAS, 323, 887

\bibitem[{{Cole} {et~al.}(2000){Cole}, {Lacey}, {Baugh}, \& {Frenk}}]{Cole2000}
{Cole}, S., {Lacey}, C.~G., {Baugh}, C.~M., \& {Frenk}, C.~S. 2000, MNRAS, 319,
  168

\bibitem[{{Colless} {et~al.}(2001)}]{2dF_colless}
{Colless}, M. {et~al.} 2001, MNRAS, 328, 1039

\bibitem[{{Colless} {et~al.}(2003)}]{2dF_final}
---. 2003, astro-ph/0306581

\bibitem[{{Conselice} {et~al.}(2003){Conselice}, {Bershady}, {Dickinson}, \&
  {Papovich}}]{C+03}
{Conselice}, C.~J., {Bershady}, M.~A., {Dickinson}, M., \& {Papovich}, C. 2003,
  AJ, 126, 1183

\bibitem[{{Couch} {et~al.}(2001){Couch}, {Balogh}, {Bower}, {Smail},
  {Glazebrook}, \& {Taylor}}]{C+01}
{Couch}, W.~J., {Balogh}, M.~L., {Bower}, R.~G., {Smail}, I., {Glazebrook}, K.,
  \& {Taylor}, M. 2001, ApJ, 549, 820

\bibitem[{{Cowie} {et~al.}(1999){Cowie}, {Songaila}, \& {Barger}}]{Cowie+99}
{Cowie}, L.~L., {Songaila}, A., \& {Barger}, A.~J. 1999, AJ, 118, 603

\bibitem[{{Coziol} {et~al.}(2000){Coziol}, {Iovino}, \& {de Carvalho}}]{Coziol}
{Coziol}, R., {Iovino}, A., \& {de Carvalho}, R.~R. 2000, AJ, 120, 47

\bibitem[{{de la Rosa} {et~al.}(2001){de la Rosa}, {de Carvalho}, \&
  {Zepf}}]{dlR01}
{de la Rosa}, I.~G., {de Carvalho}, R.~R., \& {Zepf}, S.~E. 2001, AJ, 122, 93

\bibitem[{{Di Matteo} {et~al.}(2003){Di Matteo}, {Croft}, {Springel}, \&
  {Hernquist}}]{DiMatteo03}
{Di Matteo}, T., {Croft}, R.~A.~C., {Springel}, V., \& {Hernquist}, L. 2003,
  ApJ, 593, 56

\bibitem[{{Diaferio} {et~al.}(2001){Diaferio}, {Kauffmann}, {Balogh}, {White},
  {Schade}, \& {Ellingson}}]{Diaferio}
{Diaferio}, A., {Kauffmann}, G., {Balogh}, M.~L., {White}, S.~D.~M., {Schade},
  D., \& {Ellingson}, E. 2001, MNRAS, 323, 999

\bibitem[{{Dom{\'i}nguez} {et~al.}(2001){Dom{\'i}nguez}, {Muriel}, \&
  {Lambas}}]{DML}
{Dom{\'i}nguez}, M., {Muriel}, H., \& {Lambas}, D.~G. 2001, AJ, 121, 1266

\bibitem[{{Dom{\'{\i}}nguez} {et~al.}(2002){Dom{\'{\i}}nguez}, {Zandivarez},
  {Mart{\'{\i}}nez}, {Merch{\' a}n}, {Muriel}, \& {Lambas}}]{D+02}
{Dom{\'{\i}}nguez}, M.~J., {Zandivarez}, A.~A., {Mart{\'{\i}}nez}, H.~J.,
  {Merch{\' a}n}, M.~E., {Muriel}, H., \& {Lambas}, D.~G. 2002, MNRAS, 335, 825

\bibitem[{Dressler(1980)}]{Dressler}
Dressler, A. 1980, ApJ, 236, 351

\bibitem[{{Eke} {et~al.}(2003){Eke}, {Baugh}, {Cole}, {Frenk}, {Norberg},
  {Peacock}, {et~al.}}]{Eke-groups}
{Eke}, V.~R., {Baugh}, C., {Cole}, S., {Frenk}, C.~S., {Norberg}, P.,
  {Peacock}, J.~A., {et~al.} 2003, MNRAS, in press

\bibitem[{{Ellingson} {et~al.}(2001){Ellingson}, {Lin}, {Yee}, \&
  {Carlberg}}]{Erica}
{Ellingson}, E., {Lin}, H., {Yee}, H.~K.~C., \& {Carlberg}, R.~G. 2001, ApJ,
  547, 609

\bibitem[{Fukugita {et~al.}(1995)Fukugita, Shimasaku, \& Ichikawa}]{F+95}
Fukugita, M., Shimasaku, K., \& Ichikawa, T. 1995, PASP, 107, 945

\bibitem[{{Gavazzi} {et~al.}(2002){Gavazzi}, {Boselli}, {Pedotti}, {Gallazzi},
  \& {Carrasco}}]{Gavazzi02}
{Gavazzi}, G., {Boselli}, A., {Pedotti}, P., {Gallazzi}, A., \& {Carrasco}, L.
  2002, A\&A, 396, 449

\bibitem[{{Geller} \& {Huchra}(1983)}]{Cfa3}
{Geller}, M.~J. \& {Huchra}, J.~P. 1983, ApJS, 52, 61

\bibitem[{{Girardi} {et~al.}(1998){Girardi}, {Giuricin}, {Mardirossian},
  {Mezzetti}, \& {Boschin}}]{Girardi98}
{Girardi}, M., {Giuricin}, G., {Mardirossian}, F., {Mezzetti}, M., \&
  {Boschin}, W. 1998, ApJ, 505, 74

\bibitem[{{Girardi} {et~al.}(2002){Girardi}, {Manzato}, {Mezzetti}, {Giuricin},
  \& {Limboz}}]{G+02}
{Girardi}, M., {Manzato}, P., {Mezzetti}, M., {Giuricin}, G., \& {Limboz}, F.~.
  2002, ApJ, 569, 720

\bibitem[{{Girardi} {et~al.}(2003){Girardi}, {Rigoni}, {Mardirossian}, \&
  {Mezzetti}}]{Girardi03}
{Girardi}, M., {Rigoni}, E., {Mardirossian}, F., \& {Mezzetti}, M. 2003, A\&A,
  406, 403

\bibitem[{{Gladders} \& {Yee}(2000)}]{GY00}
{Gladders}, M.~D. \& {Yee}, H.~K.~C. 2000, AJ, 120, 2148

\bibitem[{{Gnedin}(2003{\natexlab{a}})}]{Gnedin03a}
{Gnedin}, O.~Y. 2003{\natexlab{a}}, ApJ, 589, 752

\bibitem[{{Gnedin}(2003{\natexlab{b}})}]{Gnedin03b}
---. 2003{\natexlab{b}}, ApJ, 582, 141

\bibitem[{{Gomez} {et~al.}(2003){Gomez}, {Nichol}, {Miller}, {Balogh},
  {et~al.}}]{Sloan_sfr_short}
{Gomez}, P.~L., {Nichol}, R.~C., {Miller}, C.~J., {Balogh}, M.~L., {et~al.}
  2003, ApJ, 584, 210

\bibitem[{{Goto} {et~al.}(2003){Goto}, {Nichol}, {et~al.}}]{tomo-EA1}
{Goto}, T., {Nichol}, R.~C., {et~al.} 2003, PASJ, 55, 771

\bibitem[{{Granato} {et~al.}(2003){Granato}, {De Zotti}, {Silva}, {Bressan}, \&
  {Danese}}]{Granato03}
{Granato}, G.~L., {De Zotti}, G., {Silva}, L., {Bressan}, A., \& {Danese}, L.
  2003, ApJ, in press, astro-ph/0307202

\bibitem[{{Gray} \& {Moore}(2003{\natexlab{a}})}]{Gray1}
{Gray}, A. \& {Moore}, A.~W. 2003{\natexlab{a}}, in Proceedings of the Third
  SIAM International Conference on Data Mining, San Francisco, CA, USA, May
  1-3, 2003, ed. D.~Barbar{\'a} \& C.~Kamath (SIAM)

\bibitem[{{Gray} \& {Moore}(2003{\natexlab{b}})}]{Gray2}
{Gray}, A. \& {Moore}, A.~W. 2003{\natexlab{b}}, in Proceedings of the ASA
  Joint Statistical Meeting, Statistical Computing Section [CD-ROM], San
  Francisco, CA (ASA)

\bibitem[{{Gunn} \& {Gott}(1972)}]{GG}
{Gunn}, J.~E. \& {Gott}, J.~R. 1972, ApJ, 176, 1

\bibitem[{Hashimoto {et~al.}(1998)Hashimoto, Oemler, Lin, \& Tucker}]{H+98}
Hashimoto, Y., Oemler, A., Lin, H., \& Tucker, D.~L. 1998, ApJ, 499, 589

\bibitem[{{Hashimoto} \& {Oemler}(1999)}]{HO99}
{Hashimoto}, Y. \& {Oemler}, A.~J. 1999, ApJ, 510, 609

\bibitem[{{Hashimoto} \& {Oemler}(2000)}]{HO00}
---. 2000, ApJ, 530, 652

\bibitem[{{Hernquist} \& {Springel}(2003)}]{HS03}
{Hernquist}, L. \& {Springel}, V. 2003, MNRAS, 341, 1253

\bibitem[{{Hickson}(1982)}]{Hickson}
{Hickson}, P. 1982, ApJ, 255, 382

\bibitem[{{Hopkins} {et~al.}(2001){Hopkins}, {Connolly}, {Haarsma}, \&
  {Cram}}]{Hopkins01}
{Hopkins}, A.~M., {Connolly}, A.~J., {Haarsma}, D.~B., \& {Cram}, L.~E. 2001,
  AJ, 122, 288

\bibitem[{{Hopkins} {et~al.}(2003){Hopkins}, {Miller}, {Nichol},
  {et~al.}}]{Hopkins03}
{Hopkins}, A.~M., {Miller}, C.~J., {Nichol}, R.~C., {et~al.} 2003, ApJ, in
  press, astro-ph/0306621

\bibitem[{{Iglesias-P{\' a}ramo} \& {V{\'{\i}}lchez}(1999)}]{IPV}
{Iglesias-P{\' a}ramo}, J. \& {V{\'{\i}}lchez}, J.~M. 1999, ApJ, 518, 94

\bibitem[{{Jones} {et~al.}(2003){Jones}, {Ponman}, {Horton}, {Babul},
  {Ebeling}, \& {Burke}}]{Jones03}
{Jones}, L.~R., {Ponman}, T.~J., {Horton}, A., {Babul}, A., {Ebeling}, H., \&
  {Burke}, D.~J. 2003, MNRAS, 343, 627

\bibitem[{{Kauffmann} \& {Haehnelt}(2000)}]{KH}
{Kauffmann}, G. \& {Haehnelt}, M. 2000, MNRAS, 311, 576

\bibitem[{{Kauffmann} {et~al.}(2003){Kauffmann}, {Heckman}, {White},
  {et~al.}}]{Kauffmann-SDSS1_short}
{Kauffmann}, G., {Heckman}, T.~M., {White}, S.~D.~M., {et~al.} 2003, MNRAS,
  341, 54

\bibitem[{{Kelly} \& {McKay}(2003)}]{KM03}
{Kelly}, B.~C. \& {McKay}, T.~A. 2003, AJ, in press, astro-ph/0307395

\bibitem[{{Kelm} \& {Focardi}(2003)}]{KF03}
{Kelm}, B. \& {Focardi}, P. 2003, A\&A, submitted, astro-ph/0306414

\bibitem[{{Kennicutt}(1983)}]{K83}
{Kennicutt}, R.~C. 1983, ApJ, 272, 54

\bibitem[{Kennicutt(1989)}]{K89}
Kennicutt, R.~C. 1989, ApJ, 337, 761

\bibitem[{{Kennicutt}(1998)}]{Kenn_review}
{Kennicutt}, R.~C. 1998, ARA\&A, 36, 189

\bibitem[{{Kodama} {et~al.}(2001){Kodama}, {Smail}, {Nakata}, {Okamura}, \&
  {Bower}}]{Kodama_cl0939}
{Kodama}, T., {Smail}, I., {Nakata}, F., {Okamura}, S., \& {Bower}, R.~G. 2001,
  ApJL, 562, L9

\bibitem[{{Kodama} {et~al.}(2003){Kodama}, {Smail}, {Nakata}, {Okamura}, \&
  {Bower}}]{Kodama_err}
---. 2003, ApJL, 591, L169

\bibitem[{{Lambas} {et~al.}(2003){Lambas}, {Tissera}, {Alonso}, \&
  {Coldwell}}]{Lambas}
{Lambas}, D.~G., {Tissera}, P.~B., {Alonso}, M.~S., \& {Coldwell}, G. 2003,
  MNRAS, 346, 1189

\bibitem[{{Lewis} {et~al.}(2002{\natexlab{a}}){Lewis}, {Balogh}, {De Propris},
  {Couch}, {Bower}, {et~al.}}]{2dF-sfr_short}
{Lewis}, I., {Balogh}, M., {De Propris}, R., {Couch}, W., {Bower}, R., {et~al.}
  2002{\natexlab{a}}, MNRAS, 334, 673

\bibitem[{{Lewis} {et~al.}(2002{\natexlab{b}})}]{2dF_short}
{Lewis}, I.~J. {et~al.} 2002{\natexlab{b}}, MNRAS, 333, 279

\bibitem[{{Lilly} {et~al.}(1996){Lilly}, {Le Fevre}, {Hammer}, \&
  {Crampton}}]{L96}
{Lilly}, S.~J., {Le Fevre}, O., {Hammer}, F., \& {Crampton}, D. 1996, ApJL,
  460, L1

\bibitem[{{Liske} {et~al.}(2003){Liske}, {Lemon}, {Driver}, {Cross}, \&
  {Couch}}]{Liske}
{Liske}, J., {Lemon}, D.~J., {Driver}, S.~P., {Cross}, N.~J.~G., \& {Couch},
  W.~J. 2003, MNRAS, 344, 307

\bibitem[{{Madau} {et~al.}(1996){Madau}, {Ferguson}, {Dickinson}, {Giavalisco},
  {Steidel}, \& {Fruchter}}]{madau}
{Madau}, P., {Ferguson}, H.~C., {Dickinson}, M.~E., {Giavalisco}, M.,
  {Steidel}, C.~C., \& {Fruchter}, A. 1996, MNRAS, 283, 1388

\bibitem[{{Martin} \& {Kennicutt}(2001)}]{MK01}
{Martin}, C.~L. \& {Kennicutt}, R.~C. 2001, ApJ, 555, 301

\bibitem[{{Mart{\'{\i}}nez} {et~al.}(2002){Mart{\'{\i}}nez}, {Zandivarez},
  {Dom{\'{\i}}nguez}, {Merch{\' a}n}, \& {Lambas}}]{Martinez}
{Mart{\'{\i}}nez}, H.~J., {Zandivarez}, A., {Dom{\'{\i}}nguez}, M., {Merch{\'
  a}n}, M.~E., \& {Lambas}, D.~G. 2002, MNRAS, 333, L31

\bibitem[{{Mateus} \& {Sodr{\' e}}(2003)}]{MS03}
{Mateus}, A.~J. \& {Sodr{\' e}}, L.~J. 2003, MNRAS, submitted, astro-ph/0307349

\bibitem[{{McIntosh} {et~al.}(2003){McIntosh}, {Rix}, \&
  {Caldwell}}]{McIntosh03}
{McIntosh}, D., {Rix}, H.-W., \& {Caldwell}, N. 2003, ApJ, submitted,
  astro-ph/0212427

\bibitem[{{Merch{\' a}n} \& {Zandivarez}(2002)}]{MZ2dF}
{Merch{\' a}n}, M. \& {Zandivarez}, A. 2002, MNRAS, 335, 216

\bibitem[{{Merritt}(1985)}]{Merritt85}
{Merritt}, D. 1985, ApJ, 289, 18

\bibitem[{{Miller} {et~al.}(2001){Miller}, {Genovese}, {Nichol}, {Wasserman},
  {Connolly}, {Reichart}, {Hopkins}, {Schneider}, \& {Moore}}]{Miller01}
{Miller}, C.~J., {Genovese}, C., {Nichol}, R.~C., {Wasserman}, L., {Connolly},
  A., {Reichart}, D., {Hopkins}, A., {Schneider}, J., \& {Moore}, A. 2001, AJ,
  122, 3492

\bibitem[{{Miller} {et~al.}(2003){Miller}, {Nichol}, {G{\' o}mez}, {Hopkins},
  \& {Bernardi}}]{Miller_agn}
{Miller}, C.~J., {Nichol}, R.~C., {G{\' o}mez}, P.~L., {Hopkins}, A.~M., \&
  {Bernardi}, M. 2003, ApJ, 597, 142

\bibitem[{{Moss} \& {Whittle}(1993)}]{MW}
{Moss}, C. \& {Whittle}, M. 1993, ApJL, 407, L17

\bibitem[{{Moss} \& {Whittle}(2000)}]{MW00}
---. 2000, MNRAS, 317, 667

\bibitem[{{Mulchaey} \& {Zabludoff}(1998)}]{MZ98}
{Mulchaey}, J.~S. \& {Zabludoff}, A.~I. 1998, ApJ, 496, 73

\bibitem[{{Mulchaey} \& {Zabludoff}(1999)}]{MZ99}
---. 1999, ApJ, 514, 133

\bibitem[{{Nishiura} {et~al.}(2000){Nishiura}, {Shimada}, {Ohyama}, {Murayama},
  \& {Taniguchi}}]{N+00}
{Nishiura}, S., {Shimada}, M., {Ohyama}, Y., {Murayama}, T., \& {Taniguchi}, Y.
  2000, AJ, 120, 1691

\bibitem[{{Norberg} {et~al.}(2002)}]{2dF-lf2_short}
{Norberg}, P. {et~al.} 2002, MNRAS, 336, 907

\bibitem[{{O'Hely}(2000)}]{OHely}
{O'Hely}, E. 2000, PhD thesis, University of New South Wales

\bibitem[{{Patton} {et~al.}(2002){Patton}, {Pritchet}, {Carlberg},
  {et~al.}}]{Patton}
{Patton}, D.~R., {Pritchet}, C.~J., {Carlberg}, R.~G., {et~al.} 2002, ApJ, 565,
  208

\bibitem[{{Pimbblet} {et~al.}(2002){Pimbblet}, {Smail}, {Kodama}, {Couch},
  {Edge}, {Zabludoff}, \& {O'Hely}}]{kap2}
{Pimbblet}, K.~A., {Smail}, I., {Kodama}, T., {Couch}, W.~J., {Edge}, A.~C.,
  {Zabludoff}, A.~I., \& {O'Hely}, E. 2002, MNRAS, 331, 333

\bibitem[{{Poggianti} {et~al.}(2003){Poggianti}, {Bridges}, {Komiyama}, {Yagi},
  {Carter}, {Mobasher}, {Okamura}, \& {Kashikawa}}]{PoggEA03}
{Poggianti}, B.~M., {Bridges}, T.~J., {Komiyama}, Y., {Yagi}, M., {Carter}, D.,
  {Mobasher}, B., {Okamura}, S., \& {Kashikawa}, N. 2003, ApJ, in press,
  astro-ph/0309449

\bibitem[{{Poggianti} {et~al.}(1999){Poggianti}, {Smail}, {Dressler}, {Couch},
  {Barger}, {Butcher}, {Ellis}, \& {Oemler}}]{P+99}
{Poggianti}, B.~M., {Smail}, I., {Dressler}, A., {Couch}, W.~J., {Barger},
  A.~J., {Butcher}, H., {Ellis}, R.~S., \& {Oemler}, A.~J. 1999, ApJ, 518, 576

\bibitem[{{Ponman} \& {Bertram}(1993)}]{PBnat}
{Ponman}, T.~J. \& {Bertram}, D. 1993, Nature, 363, 51

\bibitem[{{Postman} \& {Geller}(1984)}]{PG84}
{Postman}, M. \& {Geller}, M.~J. 1984, ApJ, 281, 95

\bibitem[{{Quilis} {et~al.}(2000){Quilis}, {Moore}, \& {Bower}}]{QMB}
{Quilis}, V., {Moore}, B., \& {Bower}, R. 2000, Science, 288, 1617

\bibitem[{{Rocha-Pinto} {et~al.}(2000){Rocha-Pinto}, {Scalo}, {Maciel}, \&
  {Flynn}}]{R-P}
{Rocha-Pinto}, H.~J., {Scalo}, J., {Maciel}, W.~J., \& {Flynn}, C. 2000, A\&A,
  358, 869

\bibitem[{{Rubin} {et~al.}(1991){Rubin}, {Hunter}, \& {Ford}}]{RHF}
{Rubin}, V.~C., {Hunter}, D.~A., \& {Ford}, W.~K.~J. 1991, ApJS, 76, 153

\bibitem[{{Shioya} {et~al.}(2002){Shioya}, {Bekki}, {Couch}, \& {De
  Propris}}]{S+02}
{Shioya}, Y., {Bekki}, K., {Couch}, W.~J., \& {De Propris}, R. 2002, ApJ, 565,
  223

\bibitem[{{Silverman}(1986)}]{Silverman}
{Silverman}, B.~W. 1986, Density estimation for statistics and data analysis
  (Chapman-Hall: New York), 61--66

\bibitem[{{Smith} {et~al.}(2003){Smith}, {Nowak}, {Donahue}, \&
  {Stocke}}]{SNDS}
{Smith}, B.~J., {Nowak}, M., {Donahue}, M., \& {Stocke}, J. 2003, AJ, 126, 1763

\bibitem[{{Solanes} {et~al.}(2001){Solanes}, {Manrique}, {Garc{\' i}a-G{\'
  o}mez}, {Gonz{\' a}lez-Casado}, {Giovanelli}, \& {Haynes}}]{Solanes01}
{Solanes}, J., {Manrique}, A., {Garc{\' i}a-G{\' o}mez}, C., {Gonz{\'
  a}lez-Casado}, G., {Giovanelli}, R., \& {Haynes}, M.~P. 2001, ApJ, 548, 97

\bibitem[{{Strateva} {et~al.}(2001)}]{Strateva01_short}
{Strateva}, I. {et~al.} 2001, AJ, 122, 1861

\bibitem[{{Strauss} {et~al.}(2002)}]{Strauss02_short}
{Strauss}, M.~A. {et~al.} 2002, AJ, 124, 1810

\bibitem[{{Tran} {et~al.}(2001){Tran}, {Simard}, {Zabludoff}, \&
  {Mulchaey}}]{Tran}
{Tran}, K.~H., {Simard}, L., {Zabludoff}, A.~I., \& {Mulchaey}, J.~S. 2001,
  ApJ, 549, 172

\bibitem[{{Treu} {et~al.}(2003){Treu}, {Ellis}, {Kneib}, {Dressler}, {Smail},
  {Czoske}, {Oemler}, \& {Natarajan}}]{Treu03}
{Treu}, T., {Ellis}, R.~S., {Kneib}, J., {Dressler}, A., {Smail}, I., {Czoske},
  O., {Oemler}, A., \& {Natarajan}, P. 2003, ApJ, 591, 53

\bibitem[{{van Gorkom}(2003)}]{vanGorkom03}
{van Gorkom}, J.~H. 2003, in Carnegie Observatories Astrophysics Series, Vol.
  3, Clusters of Galaxies: Probes of Cosmological Structure and Galaxy
  Evolution, ed. A.~D. J.S.~Mulchaey \& A.~Omeler, Vol. astro-ph/0308209

\bibitem[{{Veilleux} {et~al.}(1999){Veilleux}, {Bland-Hawthorn}, {Cecil},
  {Tully}, \& {Miller}}]{Veilleux}
{Veilleux}, S., {Bland-Hawthorn}, J., {Cecil}, G., {Tully}, R.~B., \& {Miller},
  S.~T. 1999, ApJ, 520, 111

\bibitem[{{Verdes-Montenegro} {et~al.}(2001){Verdes-Montenegro}, {Yun},
  {Williams}, {Huchtmeier}, {Del Olmo}, \& {Perea}}]{VM01}
{Verdes-Montenegro}, L., {Yun}, M.~S., {Williams}, B.~A., {Huchtmeier}, W.~K.,
  {Del Olmo}, A., \& {Perea}, J. 2001, A\&A, 377, 812

\bibitem[{{Vollmer} {et~al.}(2000){Vollmer}, {Marcelin}, {Amram}, {Balkowski},
  {Cayatte}, \& {Garrido}}]{Vollmer00}
{Vollmer}, B., {Marcelin}, M., {Amram}, P., {Balkowski}, C., {Cayatte}, V., \&
  {Garrido}, O. 2000, A\&A, 364, 532

\bibitem[{{Wainer} \& {Thissen}(1976)}]{WT76}
{Wainer}, H. \& {Thissen}, D. 1976, Psychometrika, 41, 9

\bibitem[{{Wasserman} {et~al.}(2001){Wasserman}, {Miller}, {Nichol},
  {Genovese}, {Woncheol}, {Connolly}, {Moore}, {Schneider},
  {et~al.}}]{Wasserman}
{Wasserman}, L., {Miller}, C., {Nichol}, R.~C., {Genovese}, C., {Woncheol}, J.,
  {Connolly}, A., {Moore}, A., {Schneider}, J., {et~al.} 2001, astro-ph/0112050

\bibitem[{{Wilson} {et~al.}(2002){Wilson}, {Cowie}, {Barger}, \&
  {Burke}}]{Wilson+02}
{Wilson}, G., {Cowie}, L.~L., {Barger}, A.~J., \& {Burke}, D.~J. 2002, AJ, 124,
  1258

\bibitem[{{Wong} \& {Blitz}(2002)}]{WB02}
{Wong}, T. \& {Blitz}, L. 2002, ApJ, 569, 157

\bibitem[{{York} {et~al.}(2000)}]{SDSS_tech_short}
{York}, D.~G. {et~al.} 2000, AJ, 120, 1579

\bibitem[{{Zabludoff} \& {Mulchaey}(1998)}]{ZM98}
{Zabludoff}, A.~I. \& {Mulchaey}, J.~S. 1998, ApJ, 496, 39

\bibitem[{{Zabludoff} \& {Mulchaey}(2000)}]{ZM00}
---. 2000, ApJ, 539, 136

\bibitem[{{Zabludoff} {et~al.}(1996){Zabludoff}, {Zaritsky}, {Lin}, {Tucker},
  {Hashimoto}, {Shectman}, {Oemler}, \& {Kirshner}}]{Z+96}
{Zabludoff}, A.~I., {Zaritsky}, D., {Lin}, H., {Tucker}, D., {Hashimoto}, Y.,
  {Shectman}, S.~A., {Oemler}, A., \& {Kirshner}, R.~P. 1996, ApJ, 466, 104

\end{thebibliography}

\appendix
\section{Aperture Effects}\label{sec-apeffects}
\begin{figure}
\leavevmode \epsfysize=8cm \epsfbox{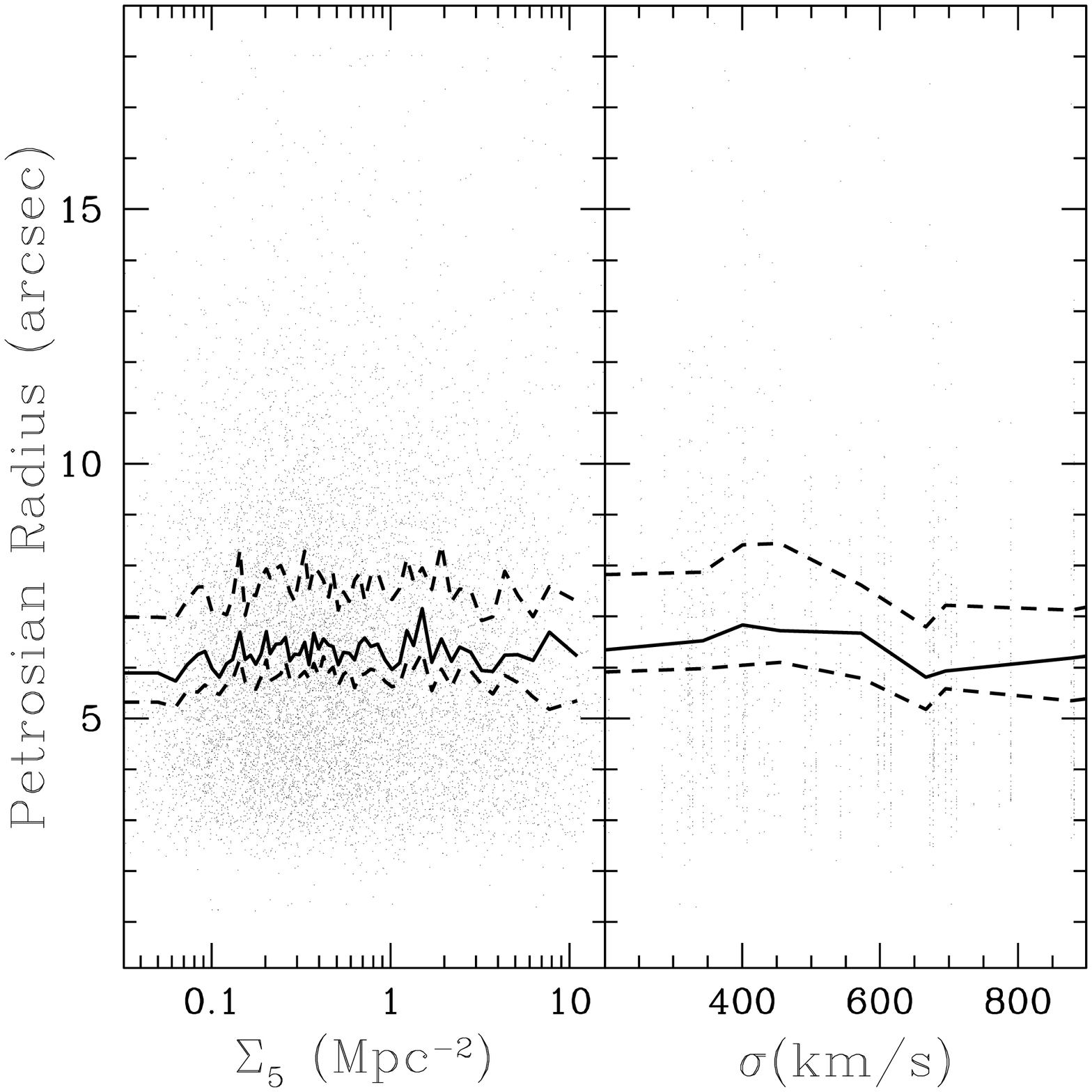}
\caption{The Petrosian radius is shown as a function of local,
  projected density $\Sigma_5$ {\it (left panel)} and of group velocity
  dispersion {\it (right panel)}, for galaxies in the SDSS sample.  The
  {\it dashed lines} show the median and \75th\ percentile of the
  distribution, in running bins each containing 500 galaxies; the {\it
    solid line} represents the mean.
\label{fig-apeff}}
\end{figure}
Since the spectra are obtained from 2\arcsec\ (2dFGRS) or 3\arcsec\
(SDSS) fibres, they do not represent the light from the whole galaxy.
In particular,  our measurements of H$\alpha$ flux will underestimate
the flux from the galaxy as a whole; we refer to this as the aperture
bias.  Aperture bias will not impact on
our results if the average galaxy size does not correlate strongly with
environment. We test this in Fig.~\ref{fig-apeff}, where we plot the SDSS
galaxy Petrosian radius as a function of $\Sigma_5$ and of group
velocity dispersion.  Note that \citet{Hopkins03} find the amount of
emission line flux lost due to aperture
bias in the SDSS is a factor $\sim 2$ for galaxies with radius $\sim 3$\arcsec,
increasing to a factor $\sim 10$ for galaxies with radius $>12$\arcsec.
However, there is no significant trend of galaxy size with
environment.  Therefore, although the absolute values of the \haew\ may
be affected\footnote{Note that \ewha\ is not necessarily
  underestimated, as it is a relative quantity.  The effect of aperture
  bias depends on the spatial distribution of star formation; if it is
  uniform, \haew\ is not affected by this bias.} by aperture bias, the relative trends as a function of
environment are secure.  This conclusion requires the further
assumption that the spatial distribution of star formation in a galaxy
depends only on its SFR, and not its environment.  For example, if star
formation in cluster galaxies is more concentrated than for galaxies in
the field with similar SFRs \citep[e.g.][]{MW,MW00}, the aperture
corrections for cluster galaxies should be smaller.

\section{Detailed comparison of the catalogues}\label{sec-cfapp}
Here we compare the 2dFGRS  and SDSS samples in detail.
The most important difference between them for our purposes is that the 2dFGRS  is selected from $b_{\rm J}$
photometric magnitudes, while the SDSS is selected from digital
$r$ photometry.  The SDSS survey is somewhat shallower, and our
volume-limited sample is complete to $M_r=-20.6$.  From the tables of \citet{F+95}, an
Sab galaxy at $z=0.06$ has a colour ($b_{\rm J}-r$)=1.13, so we adopt
a magnitude limit of $M_b=-19.5$ for the 2dFGRS  sample; the survey is actually
complete to 0.5 magnitudes fainter than this.

\begin{figure}
\leavevmode \epsfysize=8cm \epsfbox{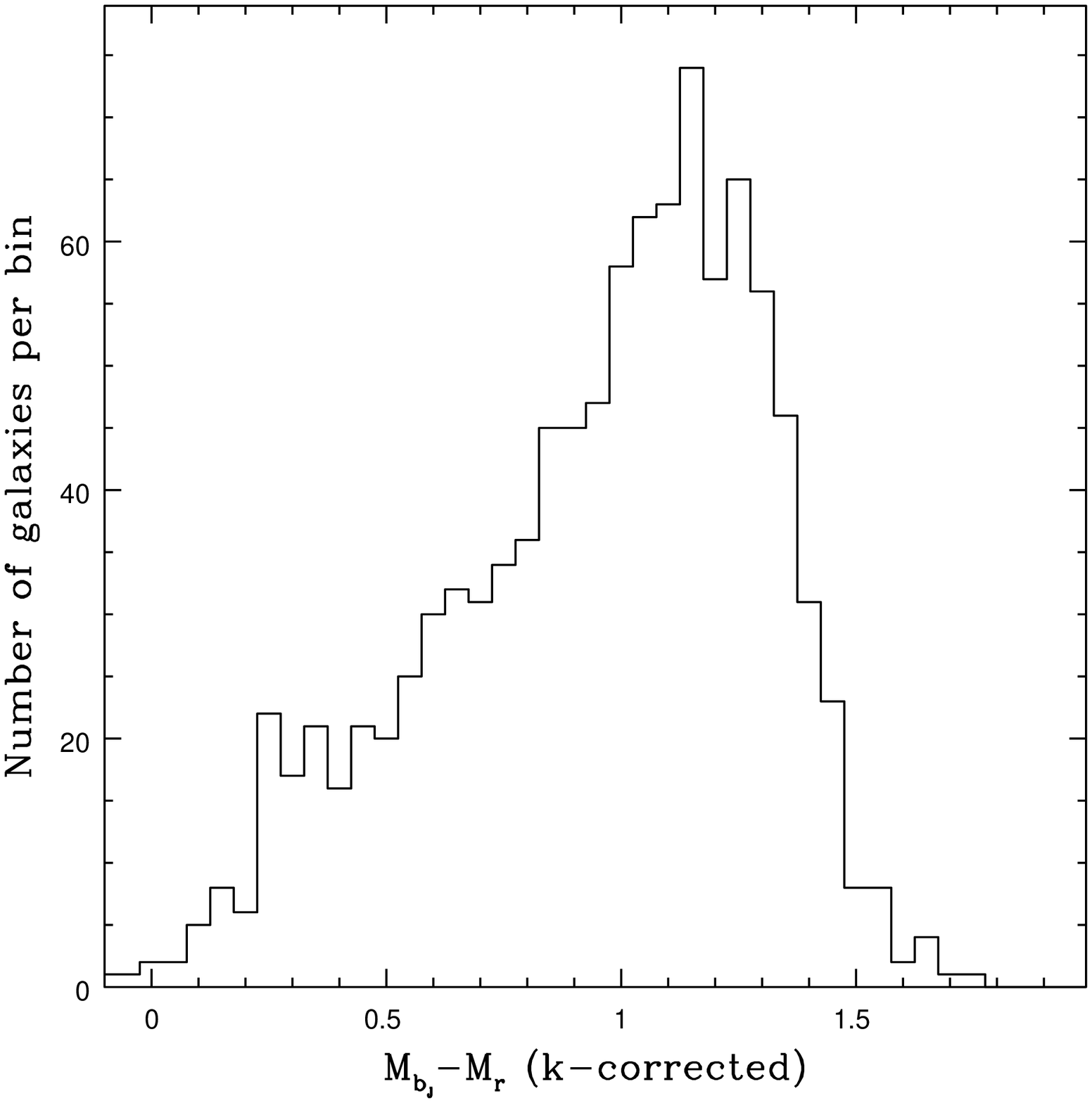}
\caption{The distribution of the differences between the absolute
  magnitudes of 1460 galaxies in our sample that are identified in both
  SDSS and 2dFGRS  catalogues. 
\label{fig-compmag}}
\end{figure}

In the area of sky covering approximately 155$<\alpha<$220 and
$-3<\delta<1.5$ degrees, the 2dFGRS  and SDSS surveys overlap.  We can
use this overlap region to directly compare the galaxy and group
catalogues.  
In our volume-limited subsample, we find 1029 galaxies in common
between the two surveys (within 2\arcsec\ and $100\kms$ of one
another). The distribution of the difference between the absolute 2dFGRS  $b_{\rm J}$ magnitude and the
absolute SDSS $r$ magnitude are shown in Fig.~\ref{fig-compmag}.
The average colour is $M_{b_{\rm J}}-M_r\sim0.95$,  similar to the colour
we expected (see above).  Thus we can be confident that our magnitude
limits are matched as closely as possible.  However, the 1$\sigma$
standard deviation in colour is $\sim 0.35$; thus there will be
large numbers of galaxies at the faint limit of one catalogue that will
not be included in the other.

\begin{figure}
\leavevmode \epsfysize=8cm \epsfbox{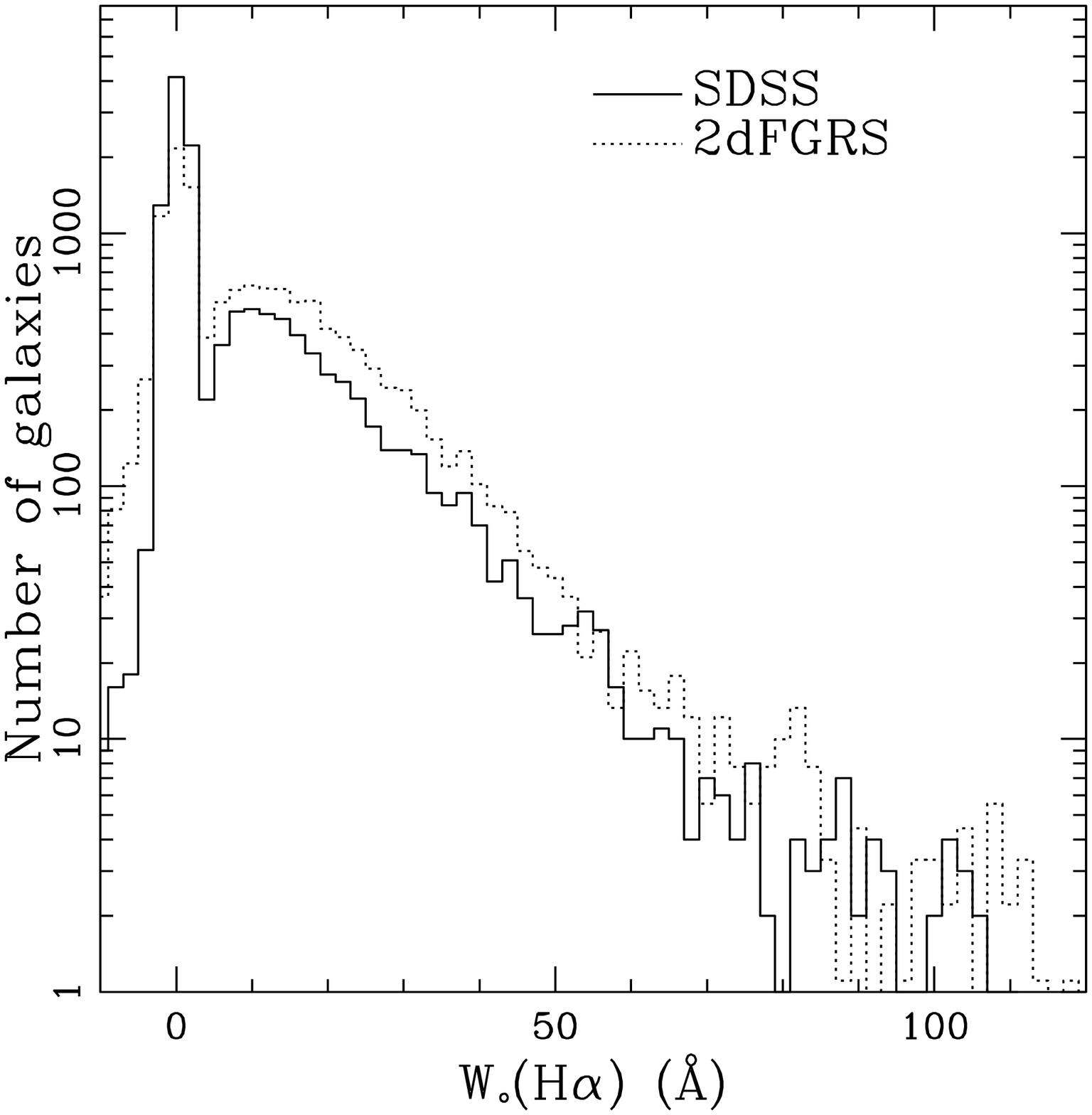}
\caption{The distribution of H$\alpha$ equivalent width in the two
  samples.  A 1 \AA\ correction for underlying stellar absorption has
  been made in both cases.  The 2dFGRS  distribution has been
  renormalised to match the number of galaxies in the SDSS, to
  facilitate the comparison.
\label{fig-hadist}}
\end{figure}

The consequences of this are shown in Fig.~\ref{fig-hadist}, where we
compare the distributions of \ewha\ in the two samples.
Recall
that an absorption correction of 1\AA\ has been applied to all measurements (see Section~\ref{sec-data}).
The two distributions are very
similar; however, there are relatively more galaxies with substantial
H$\alpha$ emission in the 2dFGRS  sample.  The mean
equivalent width is 8.1\AA\ in the SDSS, and 11.9\AA\ in the 2dFGRS; a
similar difference is seen in the \75th\ percentiles, which are also
$\sim 30$\% larger in the 2dFGRS  sample.
The larger errors on the 2dFGRS  spectra also result in a 
larger population of galaxies with \ewha$<$0 \AA, at the expense of
galaxies in the peak (with \ewha=0 \AA).

\begin{figure}
\leavevmode \epsfysize=8cm \epsfbox{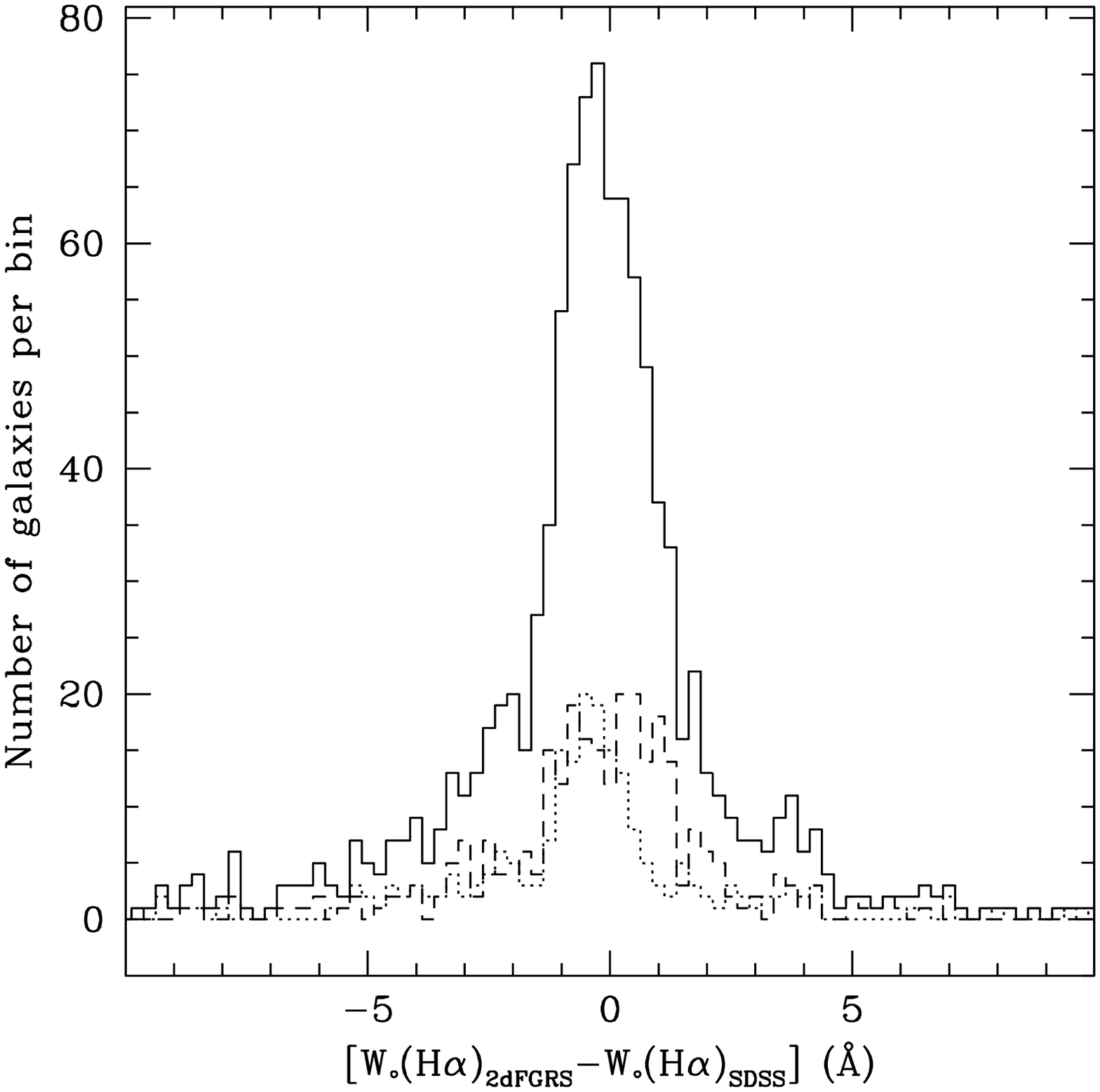}
\caption{The difference between \ewha\ measured in the 2dFGRS  and SDSS
  samples, for 1029 galaxies in common.  Thirty-three galaxies lie
  outside the bounds of this diagram.  The {\it dotted} and {\it
    dashed} lines show the same distribution, but restricted to
  galaxies at $z<0.075$ and $z>0.085$, respectively.
\label{fig-compha}}
\end{figure}

The difference in Fig.~\ref{fig-hadist} is due entirely to the blue
selection of the 2dFGRS.  In Fig.~\ref{fig-compha} we directly compare
the \ewha\ for galaxies in common between the two surveys.
Most (67\%) of the measurements are within 2\AA\ of
one another.  In the mean, the SDSS measurement of \ewha\ is 0.6\AA\ larger than
that of the 2dFGRS.   This difference is too small (and in the wrong
sense) to account for the difference in Fig.~\ref{fig-hadist}.  We also show the
comparison for galaxies at $z<0.075$ and $z>0.085$; there is no
discernible difference between them.  Since the 2dFGRS and SDSS use
different size fibres, this gives added confidence that aperture
effects do not have a large influence on our results.

\begin{figure}
\leavevmode \epsfysize=8cm \epsfbox{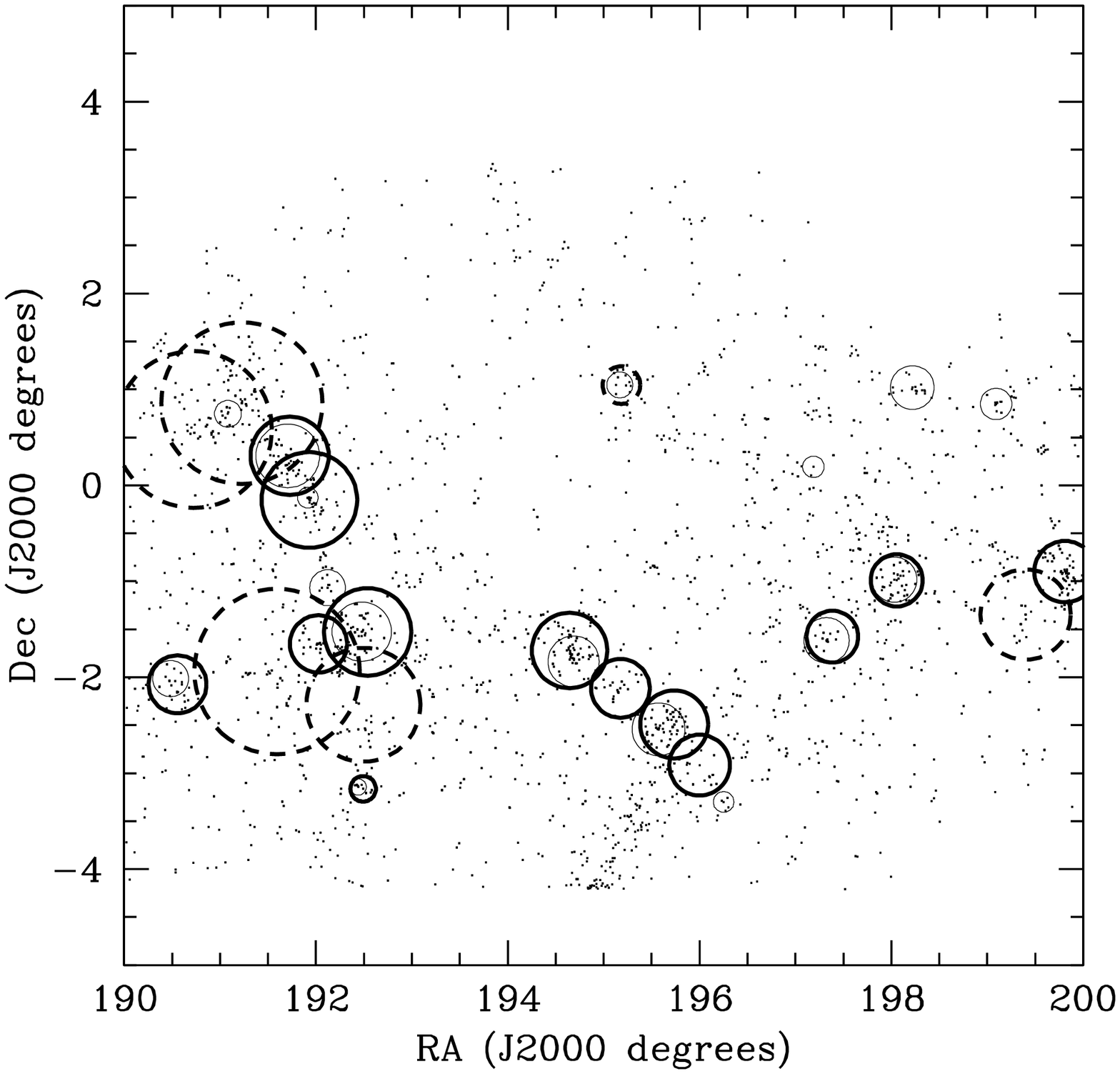}
\caption{Galaxies are shown in a 25 square degree region of space where
  the 2dFGRS and SDSS  overlap. 
Galaxy groups in the 2dFGRS sample, with $\sigma>200\kms$ and
  at least ten bright members, are shown as {\it thin circles}, with a radius
  corresponding to the virial radius (proportional to velocity dispersion).  The {\it
    thick circles} are SDSS groups; {\it dashed circles} have
  significant substructure and poorly-determined velocity dispersions.
\label{fig-cfgroups}}
\end{figure}

We can also make a direct
comparison between the group catalogues.
In Fig.~\ref{fig-cfgroups} we show the positions of 
groups in both surveys, within a $10^\circ\times7^\circ$ region.  In both cases we
show all groups within $0.05<z<0.095$ and with $\sigma>200\kms$ and
at least ten members above our luminosity limit.
For the SDSS we show all the groups satisfying these criteria; however, in this paper we
have only used those with well-determined velocity dispersions, shown
as the solid circles.  Within the full overlap region 
there are 29 such 2dFGRS  groups, compared with 24 SDSS
groups; 16 of the latter have well-determined
velocity dispersions.   The radii of the circles represents the cluster
virial radius, estimated from
the velocity dispersion as $R_{\rm vir}= 3.5 \sigma (1+z)^{-1.5}$
\citep{Girardi98}.  The correspondence between the two group catalogues
is remarkably good.  

\section{Density Estimators}\label{sec-app2}
In this Appendix, we consider two different density estimators in turn.  The
first is a projected surface density, based on counting the number of
nearest neighbours.  The second is a three-dimensional, fixed-scale
estimator derived by kernel density estimation.  

\subsection{Nearest-neighbour approach}\label{sec-NN}
The most common method is to measure the distance to the $N^{th}$ nearest
neighbour, and measure the density within that distance.  Dressler's
\citeyearpar{Dressler} original prescription called for taking the area to be
a box enclosing the tenth nearest neighbour, before background
subtraction.  Others have employed similar methods, but by using a
spherical area and first subtracting foreground and background
galaxies (e.g., Papers~I and~II).  This latter difference means that densities are typically
measured in a larger volume, since the tenth nearest neighbour is
further away after background subtraction.  In general, we will define
$d_N$ as the distance to the $N^{th}$-nearest neighbour after
background subtraction; then the
projected density is given by $\Sigma_N=N/(\pi d_N^2)$.

Although this is a two-dimensional estimate, we wish to use the
redshift information to remove the foreground and background.  In some cases, this has
been done by removing galaxies more than 3$\sigma$ from the cluster
\citep[e.g. Paper~I; ][]{lowlx-spectra_short},
where $\sigma$ is the cluster velocity dispersion.  However, this is
only reliable if $\sigma$ is well known, and the velocity distribution
is Gaussian.  We will therefore choose a fixed velocity interval of
$\pm 1000\kms$ within which to compute the local density.  This
allows us to include most of the galaxies in systems with large
velocity dispersions, while still keeping contamination low.  
We choose $N=5$ to be similar to that defined by \citet{Dressler}, who
used $N=10$ {\it before} background subtraction.

\subsubsection{Projection effects}\label{sec-proj}
\begin{figure}
\leavevmode \epsfysize=8cm \epsfbox{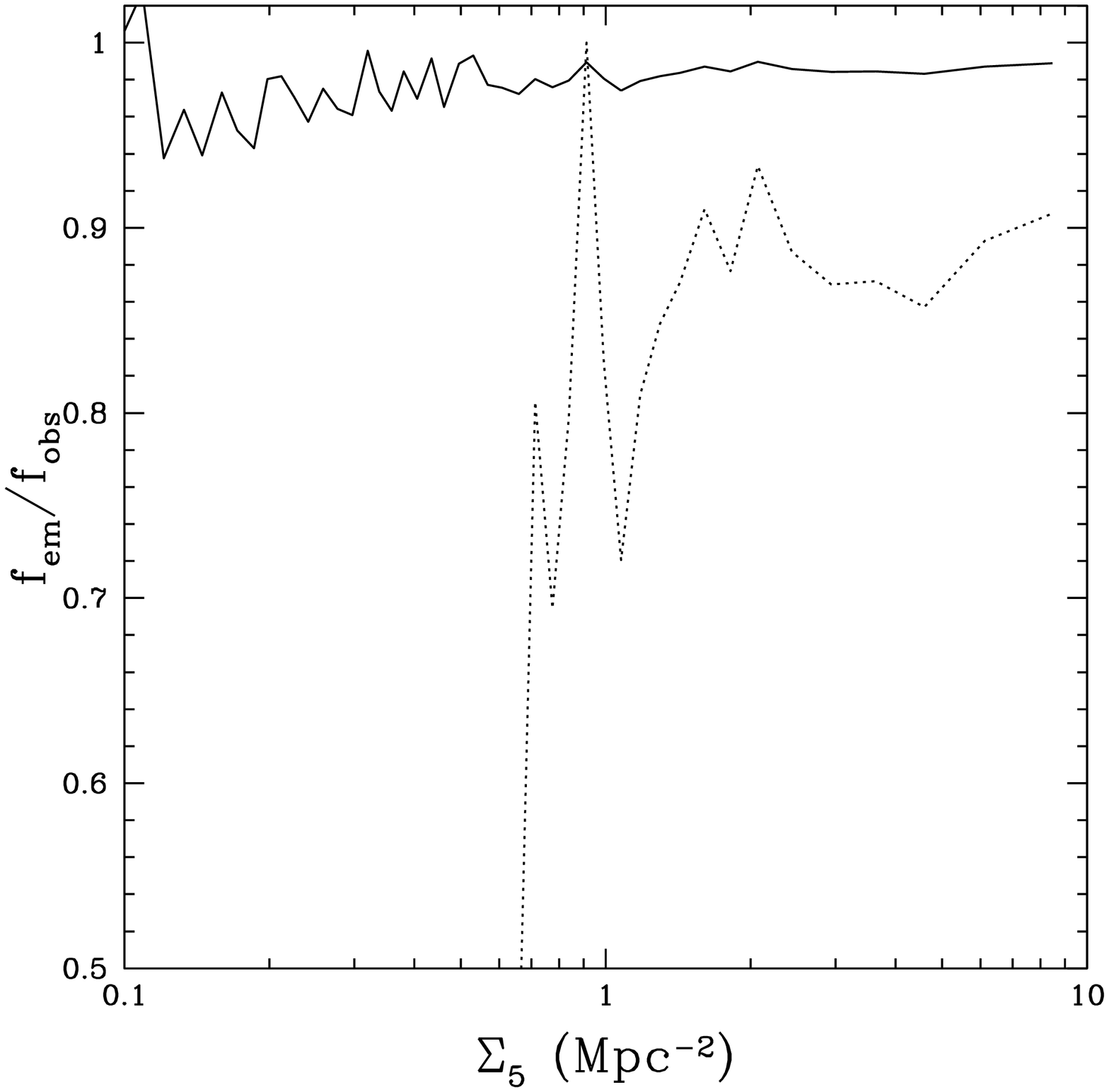}
\caption{The ratio between the true fraction of emission line galaxies
  (\ewha$>4$ \AA) and the observed fraction, which is contaminated by
  projection along the line of sight, for the 2dFGRS.  The {\it solid line} represents
  the case where the projected galaxies are at the average density of
  the Universe, while the {\it dotted line} is calculated assuming this
  density is ten times larger.
\label{fig-proj}}
\end{figure}

Our density estimator $\Sigma_5$ is a projected quantity, including all galaxies within
$\pm 1000\kms$ of the target galaxy, or $\pm 14$ Mpc for $H_0=70
\kmsmpc$.   We will use the 2dFGRS as an example (an analogous
calculation can be done for the SDSS) and estimate
the contamination at a given measured density $\Sigma_5$ by
assuming background galaxies are distributed at the mean galaxy density.
From the galaxy luminosity function of
\citet{2dF-lf2_short}, we calculate that the number of galaxies within
a projected cylinder of radius $d$ is $\sim0.17d^2$.
If we take $d_5$ to be the distance to the
fifth-nearest neighbour, then the fraction of projected galaxies within
$d_5$ is given by $f_{\rm proj}=0.17/(\pi \Sigma_5)$.  This varies from 100 per cent
at $\Sigma_5=0.054$ Mpc$^{-2}$ (which therefore represents the
field density) to 5 percent by $\Sigma_5=1$ Mpc$^{-2}$.  In
environments typical of the field, the fraction of emission-line
galaxies is $\sim 70$ per cent (Fig.~\ref{fig-haden}).  Therefore, the
true, unprojected emission-line fraction, $f_{\rm em}$, is related to
the observed fraction $f_{\rm obs}$, by $f_{\rm em}=(f_{\rm
  obs}-0.7f_{\rm proj})/(1-f_{\rm proj})$.  In Figure~\ref{fig-proj} we
show how the ratio $f_{\rm em}/f_{\rm obs}$ depends on $\Sigma_5$.  The
correction is small, $\lesssim 5$ per cent at all values of $\Sigma_5$.
We also show how this ratio changes if we assume the background is ten
times more dense than the average, which might be the case in the
vicinity of clusters and groups.  However, in this case we should adopt
a smaller intrinsic emission-line fraction, because the projected
galaxies are themselves in a more dense environment; based on
Figure~\ref{fig-haden} we take this to be 60 per cent.  Even in this case, the
correction required to account for projection is $<25$ per cent for
$\Sigma_5>1$ Mpc$^{-2}$.  In reality, such a dense field projection
is probably only reasonable at the highest $\Sigma_5$, where the
correction is $\lesssim 15$ per cent.
We conclude that projection will only have a small
effect on the trends observed in Fig.~\ref{fig-haden} and similar figures.

\subsection{Fixed scale estimates: kernel density estimation}\label{sec-kde}
\begin{figure}
\leavevmode \epsfysize=8cm \epsfbox{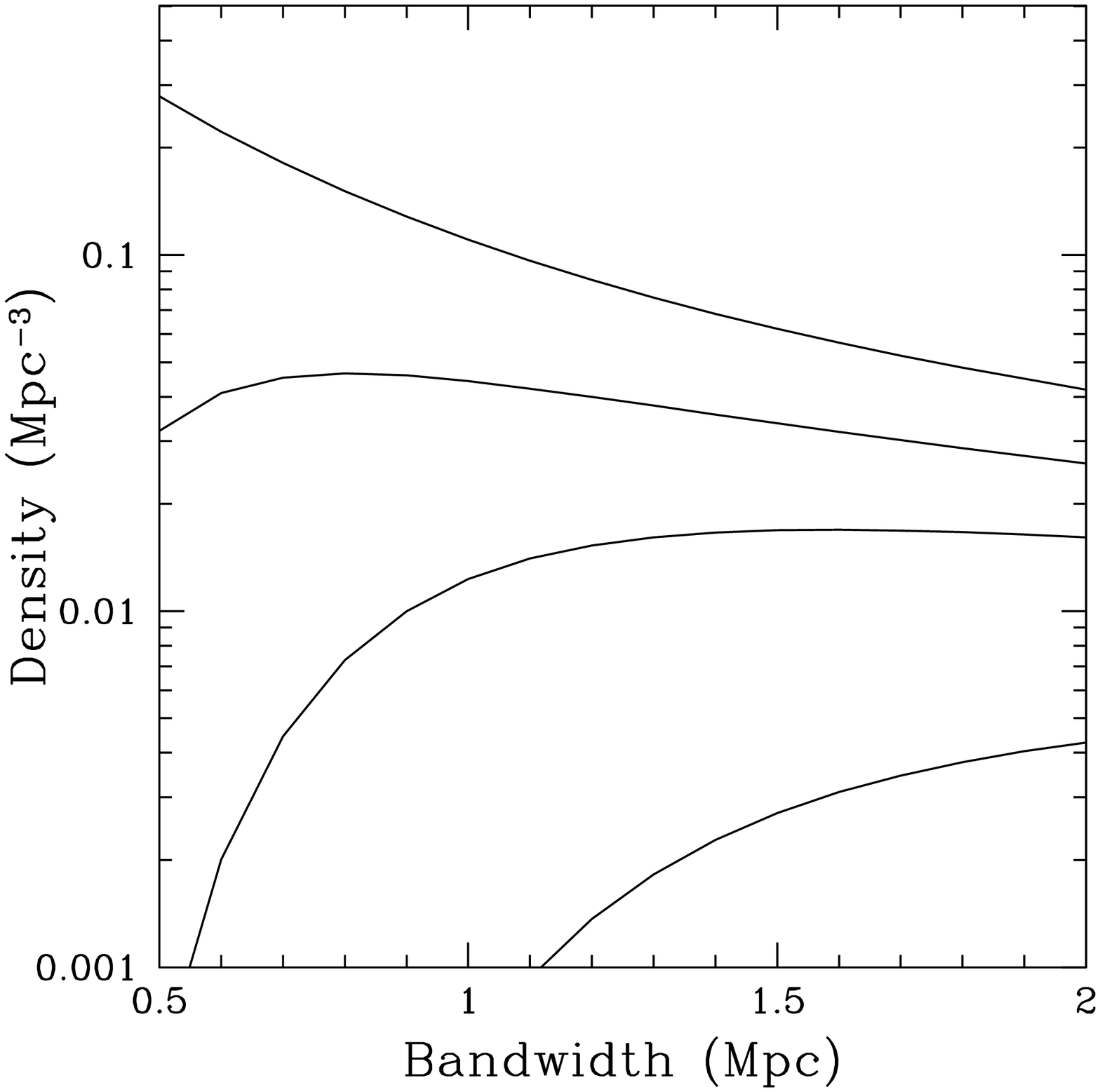}
\caption{The dependence of three-dimensional density estimate on
  bandwidth, for the combined sample.  Here, the bandwidth is the
  standard deviation of the kernel.  The sample is divided into four
  based on the quartiles of the density at $\theta=2$ Mpc.  Thus, the top
  curve shows the average density of the 25\% of galaxies which have
  the greatest density at $\theta=2$ Mpc, while the bottom curve corresponds
  to the 25\% with the lowest density at that point.
\label{fig-kdetest}}
\end{figure}

In this section, we will consider density estimates within a fixed
distance of a galaxy.  
We will use a Gaussian filtering kernel which has some weight
in the wings, so that more weight is given
to galaxies which are closer, while there is still sensitivity to
more distant structures.
A rigorous statistical analysis shows that the density estimate, for
any data distribution, is sensitive mostly to the bandwidth $\theta$ of the
smoothing kernel, and much less sensitive to the shape of the kernel \citep{Silverman}.
The strong bandwidth dependence is shown explicitly in Fig.~\ref{fig-kdetest}, where we plot the
average density as a function of $\theta$.  Four curves are shown,
corresponding to the four quartiles of the density distribution
evaluated at $\theta=2$ Mpc.  
In general, smaller bandwidths result in greater densities.
Thus, what do we choose for the optimal bandwidth?  In abstract terms, the
problem is to find an estimate $\hat{f}$ of the underlying density
distribution function $f$ which
minimises the average value of the mean squared error:
\begin{equation}\label{eqn-MSE}
MSE(f,\hat{f}) = \left< \int\left[{f(x)-\hat{f}(x)}\right]^2 dx\right>.
\end{equation}
We want to use the data itself to find the optimal bandwidth, $\theta$,
which minimises this error.  Since $\int f^2(x)dx$ does not depend on
$\theta$, this corresponds to minimising the function:
\begin{equation}
J(\theta) = \int\hat{f}^2(x)dz -2\int\hat{f}(x)f(x)dx.
\end{equation}
The method of least-squares cross-validation uses the data $X_i$ to obtain an
unbiased estimate of this function:
\begin{equation}
\hat{J}(\theta)\approx \frac{1}{\theta n^2} \sum_i\sum_j K^\ast\left(\frac{X_i-X_j}{\theta}\right)+\frac{2}{n\theta}K(0),
\end{equation}
where $K(x)$ is the kernel function, and $K^\ast=K(x)*K(x)-2K(x)$
\citep{Wasserman}.  This function can be computed using the
fast Fourier transform \citep{Silverman}, but this method is relatively
slow, and can be inaccurate due to a number of necessary
approximations.  We have implemented a new, substantially improved algorithm based on adaptive
computational geometry and a hierarchical finite-difference
approximation \citep{Gray1,Gray2}.  

\begin{figure}
\leavevmode \epsfysize=8cm \epsfbox{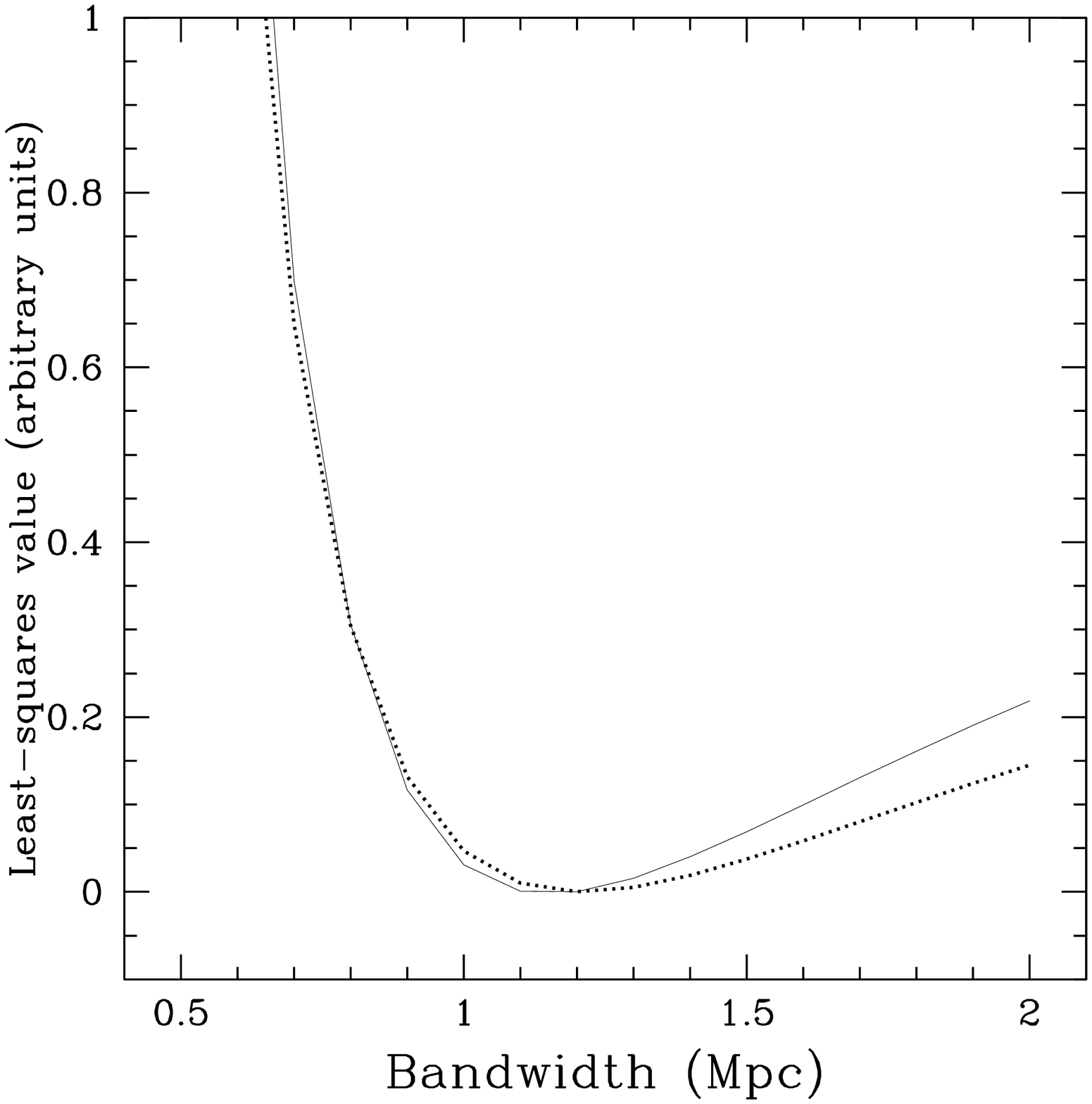}
\caption{The least-squares cross validation score (arbitrarily scaled), as a function of
  bandwidth, for the SDSS and 2dFGRS  samples ({\it solid} and {\it
    dotted} lines, respectively).  
  Both curves show a clear minimum near $\sim 1.1$ Mpc.
\label{fig-lscv}}
\end{figure}

We performed a cross-validation analysis of our
volume-limited, combined sample.  We use a Gaussian kernel for the
cross-validation, but recall that the optimal value of
$\theta$ is insensitive to the exact kernel choice.  
Here, we are computing three-dimensional densities $\rho_\theta$, assuming the
redshifts indicate line-of-sight position in a $\Lambda$ CDM model, and
ignoring the effects of redshift distortions.
Fig.~\ref{fig-lscv} shows the least-squares value as a function of bandwidth.
We see that the optimal bandwidth is $\sim 1.1$ Mpc, which we will
therefore use as our best estimate of the three-dimensional local
density.  To measure the sensitivity of galaxy properties to larger
scales, though, we will also consider densities measured on 5.5 Mpc
scales, $\rho_{5.5}$.

Note that the density estimate at a given galaxy may be arbitrarily
low, since the density at that point is measured from the surrounding
data, not including the point itself.  Densities become significantly
underestimated if the width of the Gaussian, $\theta$, is comparable to
the distance to the nearest boundary.  We therefore only
consider galaxies farther than 2$\theta$ from a survey boundary when
using these estimators.  

\begin{figure}
\leavevmode \epsfysize=15cm \epsfxsize=8cm \epsfbox{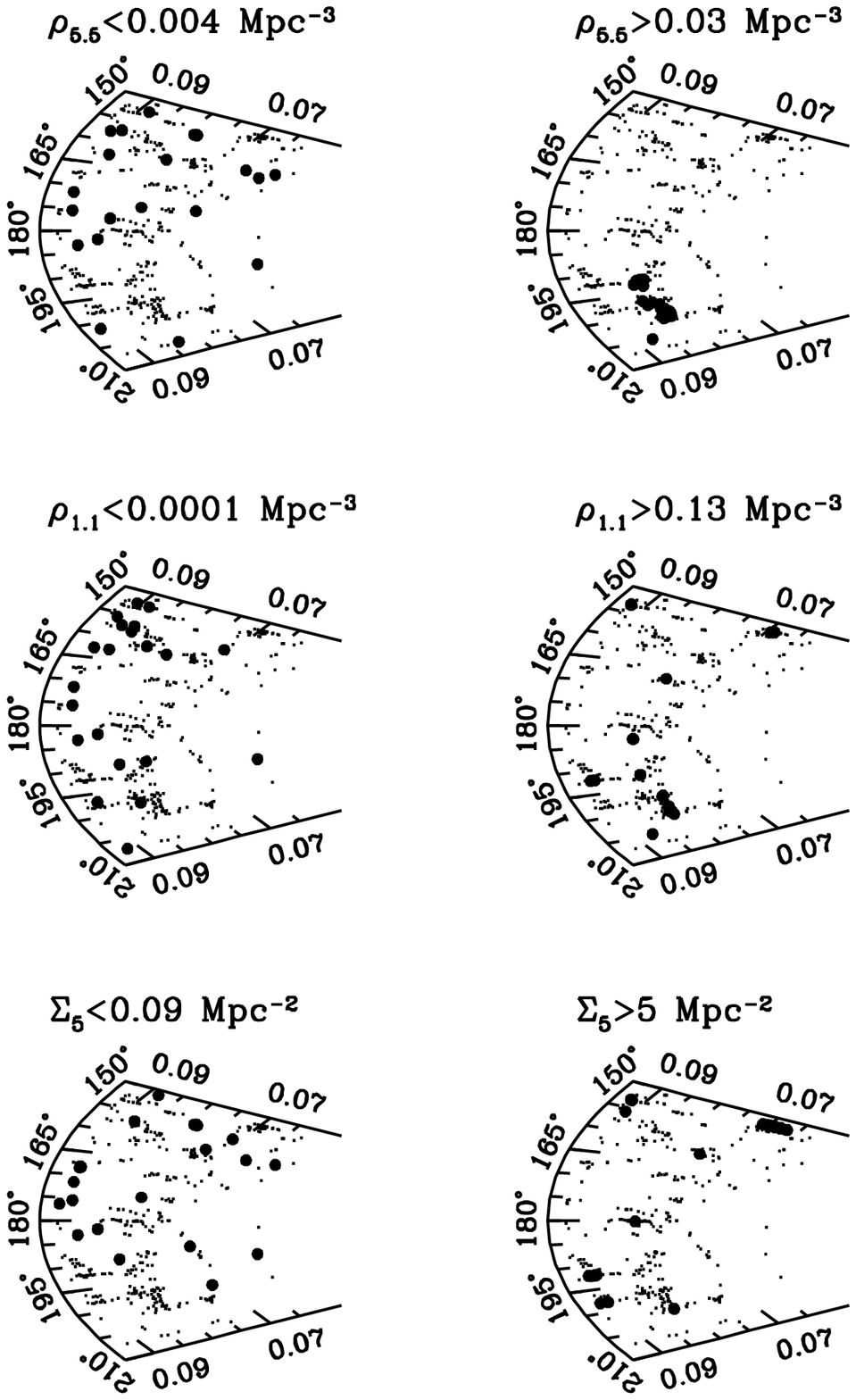}
\caption{Pie-diagrams (with redshift along the radial direction and
  right ascension as the angle) for galaxies in the 2dFGRS within 0.5 degrees of
  zero declination, and within 11 Mpc of a survey boundary.  The {\it
    filled circles} show galaxies in different density ranges, as
  indicated above each panel.
\label{fig-posden}}
\end{figure}

To see more directly what the density estimates on different scales are
measuring, we show, in Fig.~\ref{fig-posden}, pie-diagrams for a section of
the 2dFGRS.  In each diagram, we show all galaxies that lie at least 11
Mpc from the survey boundary\footnote{This motivates our choice to
  show the 2dFGRS data rather than the SDSS data.  The SDSS DR1
  catalogue is much more patchy, and there are few large,
  contiguous regions where all galaxies are at least 11 Mpc from a
  boundary.}.  Marked as thick, black circles are galaxies in different
density environments, as labelled.  For each of the three density
estimators, we show galaxies in the highest and lowest $\sim 6$ per
cent density environments.  High values of both $\Sigma_5$ and $\rho_{1.1}$
are good at finding relatively small, but dense clumps.  At low
densities, both estimators are good at finding isolated galaxies;
however, $\Sigma_5$ is unable to distinguish truly isolated galaxies
from those in low-density filaments in the plane of the sky, while
$\rho_{1.1}$ measures as low-density some cluster galaxies that are
elongated along the line of sight.  In contrast with these,
$\rho_{5.5}$ is sensitive to much larger scales.  The densest galaxies,
as measured on this scale, are only those in the massive supercluster,
while the lowest-density environments are true voids, far from
filaments or clusters.

\end{document}